\documentclass[]{aa}
\usepackage{verbatim,float}
\usepackage{times}
\usepackage{amsmath}
\usepackage{graphics}
\usepackage{epsfig}
\usepackage{psfig}

\begin{document}
\title{A wide-field photometric study of the globular cluster system of NGC 4636}
\author{B. Dirsch \inst{1}, Y. Schuberth \inst{2}, T. Richtler \inst{1}} 
\offprints{bdirsch@cepheid.cfm.udec.cl}
\institute{Universidad de Concepci\'on, Departamento de F\'{\i}sica,
           Casilla 160-C, Concepci\'on, Chile
	\and
	   Radioastronomisches Institut der Universit\"at Bonn,
	   Auf dem H\"ugel 71, 53123 Bonn, Germany
       }
\titlerunning{The Globular Cluster System of NGC 4636}

\abstract{
Previous smaller-scale studies of the globular cluster system of NGC\,4636, an
elliptical galaxy in the southern part of the Virgo cluster, have revealed an
unusually rich globular cluster system. We re-investigate the cluster system of
NGC\,4636 with wide-field Washington photometry. The globular cluster luminosity
function can be followed roughly 1\,mag beyond the turn-over magnitude found at
$\mathrm{V} = 23.31\pm0.13$ for the blue cluster sub-population. This corresponds to
a distance modulus of $(\mathrm{m}-\mathrm{M})=31.24\pm0.17$, 0.4 mag larger than the
distance determined from surface brightness fluctuations. The high specific frequency
is confirmed, yet the exact value remains uncertain because of the uncertain
distance: it varies between $5.6\pm1.2$ and $8.9\pm1.2$. The globular cluster system
has a clearly bimodal color distribution. The color peak positions show no radial
dependence and are in good agreement with the values found for other galaxies studied
in the same filter system. However, a luminosity dependence is found: brighter
clusters with an ``intermediate'' color exist. The clusters exhibit a shallow radial
distribution within 7\arcmin, represented by a power-law with an exponent of -1.4.  
Within the same radial interval, the galaxy light has a distinctly steeper profile.  
Because of the difference in the cluster and light distribution the specific
frequency increases considerably with radius. At 7\arcmin \ and 9\arcmin \ the
density profiles of the red and blue clusters, respectively, change strongly: the
power-law indices decrease to around -5 and become similar to the galaxy profile.
This steep profile indicates that we reach the outer rim of the cluster system at
approximately 11\arcmin. This interpretation is supported by the fact that in
particular the density distribution of the blue cluster population can be well fit by
the projection of a truncated power-law model with a core. This feature is seen for
the first time in a globular cluster system. While the radial distribution of the
cluster and field populations are rather different, this is not true for the
ellipticity of the system: the elongation as well as the position angle of the
cluster system agree well with the galaxy light.  We compare the radial distribution
of globular clusters with the light profiles for a sample of elliptical galaxies.  
The difference observed in NGC\,4636 is typical of an elliptical galaxy of this luminosity. 
The intrinsic specific frequency of
the blue population is considerably larger than that of the red one.
\footnote{
Tables A.1. to A.6. are only available in electronic form
at the CDS via anonymous ftp to cdsarc.u-strasbg.fr (130.79.128.5)
or via http://cdsweb.u-strasbg.fr/cgi-bin/qcat?J/A+A/
}
}

\maketitle
\section{Introduction}

The question as to what extent the study of globular cluster systems (GCSs) can
contribute to the understanding of galaxy formation has been studied by many authors
(e.g. Beasley et al. \cite{beasley02}, C\^ot\'e et al. \cite{cote02}, Forbes et al.
\cite{forbes97}, Ashman\,\&\,Zepf \cite{ashman92}) but a satisfactory answer can
still not be given. A guide to the answer could be the relation between 
global properties of GCSs and 
properties of their host galaxy. However, even when considering seemingly 
well-established statements, one faces complications when new and better observations
become available.

An example is the specific frequency $\mathrm{S}_\mathrm{N} = \mathrm{N} \cdot
10^{0.4 (\mathrm{M}_\mathrm{V}-15)}$ introduced by Harris\,\&\,van\,den\,Bergh
(\cite{harris81}), with N being the total number of globular clusters (GCs) and
$\mathrm{M}_\mathrm{V}$ the absolute visual magnitude of the host galaxy. Theory
states that central galaxies in galaxy clusters should exhibit
$\mathrm{S}_\mathrm{N}$-values about a factor 3 higher than galaxies in the field,
which typically have values of 3-4 (e.g. Harris \cite{harris01}, Elmegreen
\cite{elmegreen99}, Ashman\,\&\,Zepf \cite{ashman98}). This mainly refers to 3
central galaxies, NGC 4486 (M87) in Virgo, NGC 1399 in Fornax and NGC 3311 in Hydra
I. But recent wide-field photometry of the GCS of NGC 1399 revealed only
$\mathrm{S}_\mathrm{N} \approx 6$ (Dirsch et al. \cite{dirsch03a}), because the
galaxy's total brightness was previously underestimated (confirming an earlier
claim by Ostrov et al. \cite{ostrov98}). Although still somewhat higher than the
average for ellipticals, it turns out to be significantly lower than in previous
studies.

NGC 4636, the southernmost bright elliptical galaxy in the Virgo cluster, is a
counter example to the rule that high specific frequencies are related to central
cluster galaxies: it is a relatively isolated field galaxy and shows a peculiarly
high $\mathrm{S}_\mathrm{N}$.  In that and in other aspects, this elliptical galaxy
is intriguing. Because of its X-ray brightness, NGC\,4636 has been studied
extensively in this frequency range and these observations have also been used to
study the dark matter distribution. A surprisingly massive dark halo has been deduced
by Loewenstein\,\&\,Mushotzky (\cite{loewenstein03}): even well inside the effective
radius the dark matter constitutes a large if not the major fraction of the total
mass, which is a unique case. Its bright, diffuse X-ray halo is atypical for an
isolated galaxy and comparable in luminosity and extension to a central cluster
galaxy (e.g. Xu et al. \cite{xu02}, Jones et al. \cite{jones02}, Matsushita et al.
\cite{matsushita98}, Awaki et al. \cite{awaki94}).

NGC\,4636 also shows an unusual behavior in the far infrared. Temi et al.
(\cite{temi03}) found that its FIR luminosity between 40\,$\mu$m and 80\,$\mu$m is a
factor of 50 higher than what is expected from the dust emission produced by its
stellar body, a model that fits other elliptical galaxies quite well. Temi et al.
speculate that a recently merged, dusty, gas-rich dwarf galaxy might explain the
excess luminosity. Unfortunately, existing H\,{\sc i} observations cannot be used to
support or rule out this idea: while Knapp et al. (\cite{knapp78}), Gallagher
(\cite{gallagher78}) and Bottinelli\,\&\,Gougenheim (\cite{bottinelli78}) reported an
extended H\,{\sc i} disk with a mass of $\approx 10^9\mathrm{M}_{\sun}$,
Krishna\,Kumar\,\&\,Thonnard (\cite{krishna83}) found only an upper limit of
$10^8\mathrm{M}_{\sun}$ in NGC\,4636.

The globular cluster system of NGC 4636 has been investigated by Hanes et al.
(\cite{hanes77}) and Kissler et al. (\cite{kissler94}). They found very large
specific frequencies of $9.9$ and $7.5\pm2.5$, respectively. These values are typical
for central galaxies and not for an elliptical in a low density region such as
southern border of the Virgo cluster. With such a specific frequency NGC\,4636 would
have one of the highest $\mathrm{S}_\mathrm{N}$ values known. However, the study by
Hanes et al. was based on photographic material and Kissler et al. only used V-band
observations. It is therefore of great interest to confirm or revise the anomalous
specific frequency by new deep photometry in more than one color band. A further
objective is to determine the distance using the globular cluster luminosity function
(GCLF) and the spatial distribution of the GCs.

As illustrated for NGC\,1399 by Ostrov et al. (\cite{ostrov98}) and Dirsch et al.
(\cite{dirsch03a}), the specific frequency can easily be overestimated due to an
underestimation of the galaxy luminosity. Additionally, the specific frequency is
known to to vary radially (e.g. Ashman\,\&\,Zepf \cite{ashman98}, Larsen et al.
\cite{larsen01}, Rhode\,\&\,Zepf \cite{rhode01})). Hence it is necessary to observe
the whole radial extent of a galaxy to derive this quantity. To reliably assess the
specific frequency, it is thus crucial to obtain accurate measurements to large radii
for both, the number of GCs and the light profile.

Wide-field CCD studies exist only for five elliptical galaxies: for two GC-rich
central ellipticals: NGC\,4472 (Rhode \& Zepf \cite{rhode01}) and NGC\,1399 (Dirsch
et al. \cite{dirsch03a}) and for three more ``typical'' ellipticals: NGC\,3379,
NGC\,4406 and NGC\,4594 (Rhode\,\&\,Zepf \cite{rhode03}). The studies of the central
cluster galaxies revealed extended cluster systems that reach out to approximately
100\,kpc. While the clusters in the inner part (less than roughly 50\,kpc) have a
distinctly shallower distribution than the field population, there are indications in
both cases that field stars and clusters are similarly distributed at larger radii.

In this work we investigate the GCS of NGC\,4636 for the first time with wide-field
CCD images that cover roughly $0.5\degr\times0.5\degr$ using the 
metallicity-sensitive Washington filter system. This investigation aims to improve the
understanding of the structural parameters of the GCS and its population structure.

\section{The data}

\subsection{Observations \& reduction}

The data set consists of Washington wide-field images obtained with the MOSAIC camera
mounted at the prime focus of the CTIO 4-m Blanco telescope during 4 and 5 April
2002. We obtained four 600 sec images in R and seven 900 sec images in C.

We used the Kron-Cousins R and Washington C filters, although the genuine Washington
system uses T1 instead of R.  However, Geisler (1996) has shown that the Kron-Cousins
R filter is much more efficient than T1 and that R and T1 magnitudes are closely
related, with only a very small color term and zero-point difference. The MOSAIC
wide-field camera images a field of $36\arcmin\times 36\arcmin$. The eight SITe CCDs
have a pixel scale of 0.27\arcsec/pixel. Further information on the MOSAIC camera can
be found on the MOSAIC homepage ({\it http://www.noao.edu/kpno/mosaic/mosaic.html}).

We dithered the images to fill the gaps between the (eight) individual CCD chips. Due
to the dithering not all of the area is covered by the same number of exposures,
which restricts the usable field to $34\farcm7\times34\farcm7$.  The seeing on the
final images is 1\arcsec.

The MOSAIC data has been handled using the {\it mscred} package within IRAF. In
particular this software corrects for the variable pixel scale across the CCD which
would otherwise cause a 4\% variability of the brightness of star-like objects from
the center to the corners. The flatfielding resulted in images that had remaining
sensitivity variations $\le 2.5\%$ (peak-to-peak).

To facilitate the search for point sources the extended galaxy light was subtracted.  
This was done using a median filter with an inner radius of $9.5\arcsec$ and an outer
radius of $11\arcsec$. This size is large enough to not alter the point source
photometry which has been verified with artificial star tests described later on.

\subsection{Photometry}

The photometry has been done using the PSF fitting routine {\it allstar} within
DAOPhot II.

For the point source selection we used the $\chi$ and sharpness values from the PSF
fit and found approximately 10000 point sources. The brightest non-saturated objects
have $\mathrm{T1}\approx 18.5$ (depending slightly on the individual MOSAIC chip).

In each night 4 to 5 fields, each containing about 10 standard stars from the list of
Geisler (\cite{geisler96a}), were observed with a large coverage of airmasses
(typically from 1.0 to 1.9). It was possible to use a single transformation for both
nights, since the coefficients were indistinguishable within the uncertainties.

We derived the following relations between instrumental and standard magnitudes: 
\begin{eqnarray} 
 \begin{split} \nonumber
	\mathrm{T1} =
	&\mathrm{r}+(0.71\pm0.01)-(0.07\pm0.01)X\\
		&+(0.033\pm0.003)(\mathrm{C}-\mathrm{T1})\\
	\mathrm{(C-T1)} =&(\mathrm{c}-\mathrm{r})-(0.74\pm0.02)-(0.20\pm0.01)X\\
		&+(0.088\pm0.004)(\mathrm{C}-\mathrm{T1})
\end{split}
\end{eqnarray}

The standard deviation of the difference between instrumental and calibrated
magnitudes is 0.021 in T1 and 0.023 in C-T1.

A table containing the cluster candidates with coordinates and photometric data is
available in electronic form.

\subsection{Photometric completeness}

\begin{figure}[t]
 \centerline{\resizebox{\hsize}{!}{\includegraphics{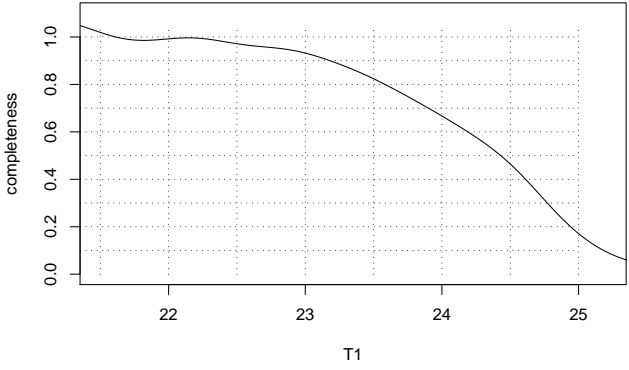}}}
 \caption{The average completeness function of the entire
	field.}
 \label{dirsch.fig2}
\end{figure}

The completeness of the data has been studied with the aid of the task {\it addstar}
within DAOPhot II, which was used to add 6666 stars to the science image. This number
was found not to increase the crowding in the field. This was done ten times to
produce ten different images. These modified images were reduced in the same way as
the original data. The final, spatially averaged completeness function for the entire
MOSAIC field in the color range $0.9<(\mathrm{C-T1})<2.5$ is plotted in
Fig.\,\ref{dirsch.fig2} for the entire MOSAIC field. The difference between the
completeness function for red and blue clusters is marginal and does not need to be
considered. However, there are strong spatial variations: close to the center of the
elliptical galaxy and in the vicinity of bright foreground stars the completeness is
lower. Throughout our study we applied completeness corrections as a function of
spatial range.

\subsection{The color-magnitude diagram}

\begin{figure}[t]
 \centerline{\resizebox{\hsize}{!}{\includegraphics{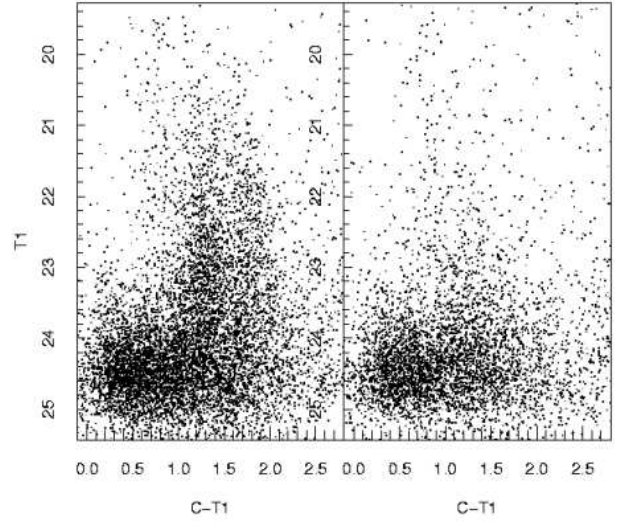}}}
 \caption{{\bf Left panel:} the color magnitude diagram (CMD) of 
	objects with a radial distance smaller than $13\farcm5$.
	{\bf Right panel:} the CMD of those objects further away than $13\farcm5$
	from NGC\,4636. We argue in Sect.\,3 that in the more
	distant sample the number of globular clusters is negligible. 
	This sample is used as background. The globular cluster population is discernible 
	within $0.9<(\mathrm{C-T1})<2.1$.}
 \label{dirsch.fig3}
\end{figure}

The color magnitude diagrams for all point sources in the MOSAIC images are plotted
in Fig.\,\ref{dirsch.fig3}. The left panel shows all objects within a galactocentric
distance of $13\farcm5$ which we consider to be the appropriate distance to separate
clusters from background objects (see Sect.\,5.1). The so defined background objects
are plotted in the right panel. The globular cluster population is discernible within
$0.9<(\mathrm{C-T1})<2.1$. The bimodal nature of the color distribution (see Sect.3)
is already visible. Background galaxies become significant at R$>23$ and begin to
dominate the CMD at R$>24$. Most foreground stars are redder than the color range
used for these CMDs.

For the foreground absorption Schlegel et al. (\cite{schlegel98}) give
$\mathrm{A}_\mathrm{B}=0.118$ and Burstein\,\&\,Heiles (\cite{burstein82})
$\mathrm{A}_\mathrm{B}=0.05$. Using the reddening law of Rieke\,\&\,Lebofsky
(\cite{rieke85}) and E$_{\mathrm{C-T1}}=1.97$\,E$_{\mathrm{B-V}}$
(Harris\,\&\,Canterna \cite{harris77}), we find for the extinction and absorption
$\mathrm{A}_\mathrm{T1}=0.08/0.04$, $\mathrm{E}_\mathrm{C-T1} = 0.06/0.03$ (Schlegel
et al./Burstein\,\&\,Heiles values), respectively. In the following we adopt the mean
values, $\mathrm{A}_\mathrm{T1}=0.06\pm0.03$ and $\mathrm{E}_\mathrm{C-T1} =
0.04\pm0.02$.

\section{Cluster color distribution}

\begin{figure}[t]
 \centerline{\resizebox{\hsize}{!}{\includegraphics{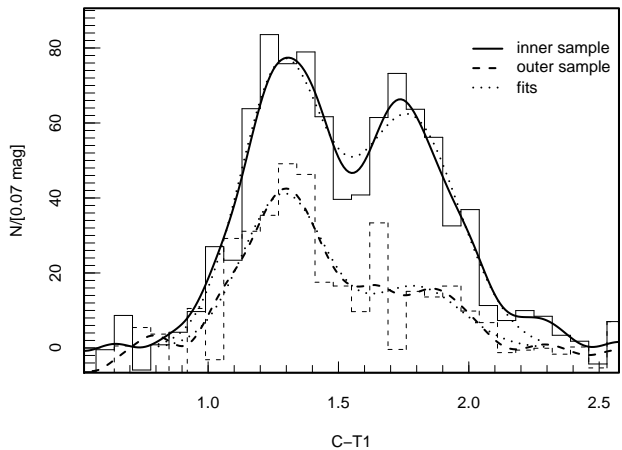}}}
 \caption{The color distribution of cluster candidates (brighter than R=24) is shown for an
	inner ($3\farcm6<r<8\farcm1$) and an outer ($8\farcm1<r<13\farcm5$) 
	sample with solid and dashed lines, respectively. 
	The sum of two Gaussians is fit to both distributions,  plotted with dotted lines.}
 \label{dirsch.fig4}
\end{figure}

The color distribution of the cluster candidates is clearly bimodal as shown in
Fig.\,\ref{dirsch.fig4}.  In this figure, as in the whole paper, the data is shown as
histograms and as an adaptive kernel filtered distribution (using an
Epanechikov kernel, Merritt\,\&\,Tremblay \cite{merritt94}). Whenever a fit is
performed, the histogram data is used. This figure also shows how the distribution
varies with radius:  the fraction of red clusters (those with
$1.55<(\mathrm{C-T1})<2.1$) is higher at smaller radii ($3\farcm6<r<8\farcm1$) than
at larger ($8\farcm1<r<13\farcm5$). A sum of two Gaussians is fitted to the color
histograms:  
$$\mathrm{a}_1\mathrm{e}^{-((\mathrm{C-T1})-\mathrm{b}_1)^2/(2\sigma_1^2)} +
\mathrm{a}_2 \mathrm{e}^{-((\mathrm{C-T1})-\mathrm{b}_2)^2/(2\sigma_2^2)}.$$ The fit
results are tabulated in Table\,\ref{tab:gaussfit}.

\begin{table*}
\begin{tabular}{ccccccc}
\hline
\hline
radial range & \multicolumn{3}{c}{blue peak} & \multicolumn{3}{c}{red peak} \\\hline
$3\farcm6<r<8\farcm1$ & $\mathrm{a}_1=84\pm4$ & $\mathrm{b}_1=1.28\pm0.02$ & $\sigma_1=0.15\pm0.02$ & 
  $\mathrm{a}_2=67\pm4$ & $\mathrm{b}_2=1.77\pm0.02$ & $\sigma_2=0.19\pm0.02$ \\
$8\farcm1<r<13\farcm5$& $\mathrm{a}_1=46\pm6$ & $\mathrm{b}_1=1.28\pm0.02$ & $\sigma_1=0.14\pm0.02$ &
  $\mathrm{a}_2=16\pm4$ & $\mathrm{b}_2=1.78\pm0.10$ & $\sigma_2=0.20\pm0.10$ \\\hline
\end{tabular}
\caption{The results of a two-Gaussian fit to the  color distribution of the GC candidates 
	brighter than R=24.}
\label{tab:gaussfit}
\end{table*}

This fit shows that the peak position and the width of the Gaussians for the two
populations remain constant with radius, despite the change of the relative heights.
The radial constancy of the peak position has also been observed in the V-I color for
a sample of elliptical galaxies by Larsen et al. (\cite{larsen01}).

In the subsequent analysis we separate the clusters into two sub-samples, a red one
($1.55<(\mathrm{C-T1})<2.1$) and a blue one ($0.9<(\mathrm{C-T1})<1.55$) divided at
the minimum of the inner sample plotted in Fig.\,\ref{dirsch.fig4}.

\begin{figure}[t]
 \centerline{\resizebox{\hsize}{!}{\includegraphics{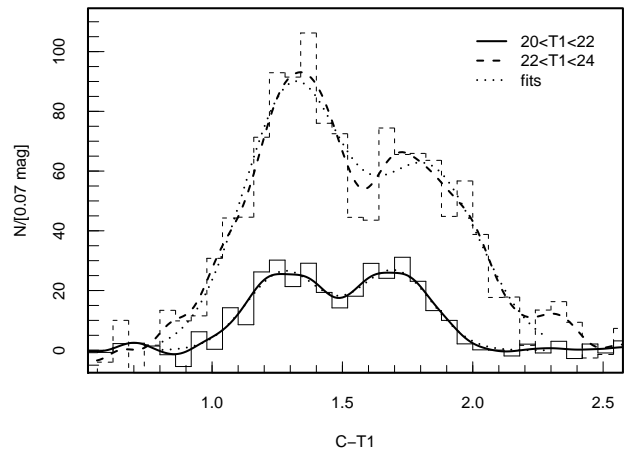}}}
 \caption{The color distribution of cluster candidates shown for a
	bright ($20<T1<22$) and a faint ($22<T1<24$) 
	sample with solid and dashed line, respectively. For both samples 
	the sum of two Gaussians is fit, shown by the dotted lines.}
 \label{dirsch.fig4_faintbright}
\end{figure}

The color distribution depends on the luminosity:  the fraction of bright red
clusters compared to blue ones is larger than the corresponding fraction for faint
clusters, indicating that the luminosity functions are different.  This is shown in
Fig.\ref{dirsch.fig4_faintbright}. The fitted peak positions are
$1.28\pm0.02/1.84\pm0.04$ for the faint sample and $1.31\pm0.04/1.69\pm0.02$ for the
bright sample. The color peak of the bright red clusters is approximately 0.1 mag
bluer than that of the fainter cluster sample. We regard this as an indication that the
concept of only two populations might not be sufficient.

\section{The luminosity function of the cluster population}

\begin{figure}[t]
 \centerline{\resizebox{\hsize}{!}{\includegraphics{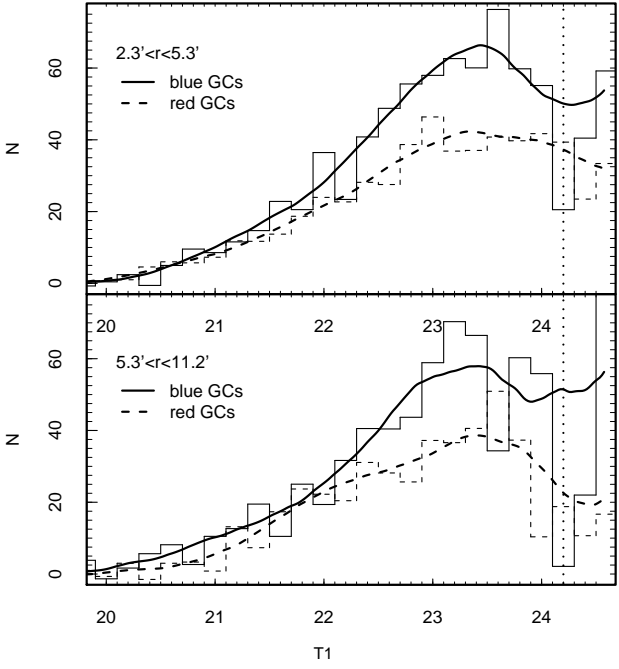}}}
 \caption{The luminosity functions for the blue ($0.9<(\mathrm{C-T1})<1.55$)	
	and the red ($1.55<(\mathrm{C-T1})<2.1$) GCs for an
	inner sample ($2\farcm3<r<5\farcm3$) in the {\bf upper panel} and 
	for an outer sample ($5\farcm3<r<11\farcm2$) in the {\bf lower panel}.
	For all samples we plotted a histogram and a kernel smoothed distribution.
	The errors of the histogram counts can be found in Table\,2 and were 
	omitted in the figure. The average 50\% completeness limit is marked by
	the vertical dotted line.}
 \label{dirsch.fig6}
\end{figure}

The LF of the cluster candidates is shown in Fig.\,\ref{dirsch.fig6} for two radial
intervals ($2\farcm3<r<5\farcm3$ and $5\farcm3<r<11\farcm2$) and for the red
($0.9<(\mathrm{C-T1})<1.55$) and the blue ($1.55<(\mathrm{C-T1})<2.1$) cluster
samples. The GCLF can be described by a Gaussian or a t5 function which has more
extended wings (Secker\,\&\,Harris \cite{secker93}). However, no systematic
differences in the turn over magnitudes (TOMs) have been reported when using these
functions (Larsen et al. \cite{larsen01}). We adopt the Gaussian model. The results
of fitting the blue, the red and the entire cluster sample are shown in the first
three columns of Table\,\ref{tab:lktfits}. Since the position of the starting bin is
arbitrary, we fitted two histograms with starting bins that are half the bin size
apart and averaged the results which always deviated by less than 0.2 mag.

We have shown in Section\,3 that the brighter clusters have a bluer red peak than the
fainter ones. This indicates that two populations are not sufficient to describe the
cluster system. We therefore divide the red cluster sample further, into an
''intermediate'' color sample ($1.55<(\mathrm{C-T1})<1.8$) and a ``very red'' sample
($1.8<(\mathrm{C-T1})<2.2$). The ''intermediate'' color sample selects the red peak
of the bright cluster population.  The results of a Gaussian fit to the data are
found in Table\,\ref{tab:lktfits}.

\begin{figure}[t]
 \centerline{\resizebox{\hsize}{!}{\includegraphics{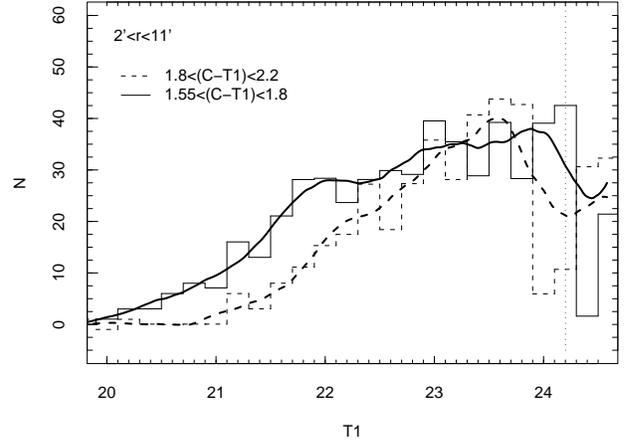}}}
 \caption{The luminosity distributions of cluster candidates at larger radii for
	a red ($1.8<(\mathrm{C-T1})<2.15$) and an ''intermediate'' color
	($1.55<(\mathrm{C-T1})<1.8$) sample are plotted 
	with solid and dashed lines, respectively.}
 \label{dirsch.fig7}
\end{figure}

While most sub-samples are found to have similar TOMs and widths (which will be
detailed shortly) the fit parameters for the intermediate color sample deviate
significantly: the TOM is about one magnitude fainter and the width a factor of two
broader. Moreover, the uncertainties are much larger. Despite the faint TOM we do not
claim that there are relatively more faint clusters. We rather consider the faint
TOMs with the enormous errors (at least a factor of 3 larger than in the other cases)
as an indication that the fit of a Gaussian is not adequate. This interpretation is
illustrated with Fig.\,\ref{dirsch.fig7} that compares the LF of the intermediate and
the very red sub-samples. This figure shows that the fraction of bright GCs is larger
for the intermediate color sample, consistent with the results using the color
distributions.

It is interesting to see this behavior in the context of the luminosity dependent
color distribution of the GCS in NGC\,1399: in this galaxy the bimodality disappears
for the brightest GCs, which means a larger fraction of ``intermediate'' colored GC
candidates. The color peak of the brightest clusters in NGC\,1399 is at
$\mathrm{C}-\mathrm{T1}=1.6\pm0.1$ while the ``intermediate'' colored sample of
NGC\,4636 is in the range $1.55<\mathrm{C}-\mathrm{T1}<1.8$, hence the majority is
redder than in NGC\,1399. A further discussion of the potential nature of these
sources is given in Sect.\,7.2.

Apart from the intermediate color sample no systematic differences in the TOM values
can be seen within the uncertainties. This is in contrast to the expected
dependence of the TOM on the metallicity of the cluster sample predicted by Ashman et
al. (\cite{ashman95}). This discrepancy is discussed in more detail in Sect.\,7.2.
The TOM that will be used from now on is the mean value of the blue
($0.9<(\mathrm{C}-\mathrm{T1})<1.55$) and the very red
($1.8<(\mathrm{C}-\mathrm{T1})<2.1$)  samples in the entire field: $\mathrm{T1} =
23.31\pm0.13$ and the width $\sigma=1.03\pm0.15$.

The TOM and the width of the fitted Gaussian are strongly correlated and we find that
$\mathrm{TOM}_\mathrm{T1}=(23.31\pm0.13)+(0.7\pm0.11)\cdot\Delta\sigma$ is a good fit
where $\Delta\sigma$ is the difference from the adopted width of 1.03. For example,
for a width of $\sigma=1.3$ which is frequently kept constant when fitting the TOM of
an elliptical galaxy in R which is comparable to T1, the TOM would be
$\mathrm{T1}=23.49\pm0.13$.

\begin{table*}
\begin{tabular}{cccccc}
\hline
\hline
radial & $(\mathrm{C}-\mathrm{T1})$ &  $(\mathrm{C}-\mathrm{T1})$ 
	& $(\mathrm{C}-\mathrm{T1})$ &  $(\mathrm{C}-\mathrm{T1})$
	& $(\mathrm{C}-\mathrm{T1})$  \\
interval & [0.9-2.1] & [0.9-1.55] & [1.55-2.1] & [1.55-1.8] & [1.8-2.15] \\\hline
$1\farcm3 - 11\farcm3$ & $23.38 \pm 0.14$ & $23.37 \pm 0.11$ & $23.41 \pm 0.21$ & $24.15 \pm 0.77$ & $23.24 \pm 0.14$ \\
		     & $1.22 \pm 0.13$  & $1.17 \pm 0.12$  & $1.20 \pm 0.20$  & $2.10 \pm 0.50$  & $0.86 \pm 0.17$   \\
$1\farcm3 - 4\farcm5$  & $23.44 \pm 0.18$ & $23.32 \pm 0.11$ & $23.63 \pm 0.27$ & $24.70 \pm 1.30$ & $23.41 \pm 0.21$ \\
		     & $1.26 \pm 0.16$  & $1.11 \pm 0.12$  & $1.44 \pm 0.23$  & $2.40 \pm 0.90$  & $1.01 \pm 0.21$  \\
$4\farcm5 - 11\farcm3$ & $23.36 \pm 0.20$ & $23.42 \pm 0.25$ & $23.29 \pm 0.25$ & $24.02 \pm 1.02$ & $23.18 \pm 0.15$  \\
		     & $1.14 \pm 0.21$  & $1.14 \pm 0.23$  & $1.14 \pm 0.27$  & $1.94 \pm 0.78$  & $0.79 \pm 0.17$  \\\hline
\end{tabular}
\caption{The LF of GC candidates with $\mathrm{T1}<24$ has been fitted with a Gaussian 
	of variable width and TOM for several sub-samples. For each radial interval two lines
	are given, the first showing the TOM, the second the width of the Gaussian ($\sigma$).}
\label{tab:lktfits}
\end{table*}

\subsection{The distance to NGC\,4636}

The absolute value of the TOM is best established in the V system (e.g.  Richtler
\cite{richtler03}, Harris \cite{harris01}, Ashman\,\&\,Zepf \cite{ashman98}).
However, the V-R color of our clusters is unknown and the only possibility is to
shift their R-TOM by a constant value. We assume a mean $\mathrm{V-T1}=0.50\pm0.05$
for our clusters (which has been derived from Galactic clusters of similar
$\mathrm{C-T1}$ color from the lists in Harris\,\&\,Canterna (\cite{harris77}) and
Harris \cite{harris96}). Thus we find for the reddening corrected TOM $\mathrm{V} =
23.76\pm0.13$

Theoretically the TOM should depend on the metallicity (Ashman et al.
\cite{ashman92}). However, in our sample no such effect is visible. For this reason
we do not attempt to correct for any metallicity dependence.  According to Richtler
(\cite{richtler03}) the TOM of the Milky Way, M31 and the mean of a (selected) sample
of elliptical galaxies with SBF measurements is the same within the uncertainties (of
the order 0.2\,mag for each method), despite the possible metallicity differences
between the GCS of these galaxies.  He gives an averaged absolute TOM of
$-7.48\pm0.11$ in V.  We hence find a distance modulus of $(\mathrm{m} -\mathrm{M}) =
31.24 \pm 0.17$.

Alternatively, we determined the R-TOM for the Milky Way using the GC compilation of
Harris (\cite{harris96}) and find $\mathrm{TOM}_\mathrm{R} = -8.0\pm0.16$. With this
value we determine a distance modulus of $(\mathrm{m} -\mathrm{M}) = 31.31\pm0.21$,
which is compatible with the above quoted value.

The obtained distance modulus agrees with that of Kissler et al. (\cite{kissler94}),
$(\mathrm{m}-\mathrm{M})=31.4\pm0.3$ which is also based on the GCLF. However, the
comparison with the distance modulus based on the SBF method is less favorable
($(\mathrm{m}-\mathrm{M})=30.83\pm0.13$, Tonry et al. \cite{tonry01}). The latter
would correspond to a $\mathrm{TOM}_\mathrm{T1}=22.87\pm0.18$.  We comment on this
difference in the discussion.

\section{Spatial distribution}

\subsection{Radial cluster distribution}

\begin{figure}[t]
 \centerline{\resizebox{\hsize}{!}{\includegraphics{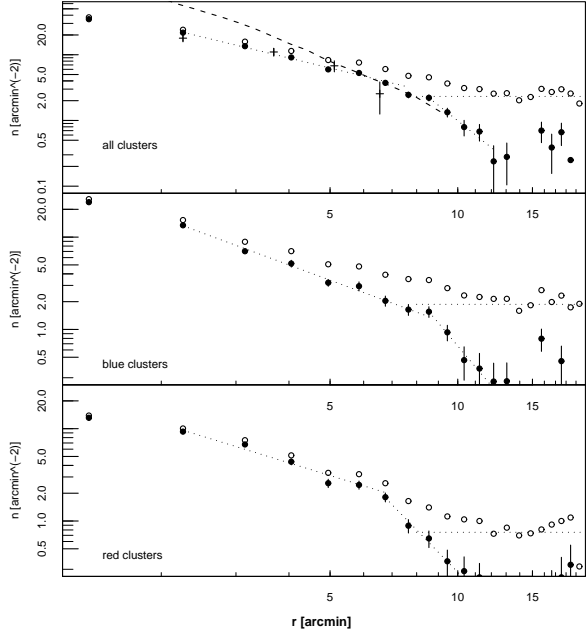}}}
 \caption{The radial distribution of all cluster candidates, the blue  
	($0.9<(\mathrm{C-T1})<1.55$) and the red ($1.55<(\mathrm{C-T1})<2.1$)
	sub-samples are shown in the three
	panels from top to bottom, respectively. The limiting magnitude is T1=24. 
	We also included the
	radial cluster density profile of Kissler et al. (\cite{kissler94})
	in the uppermost panel using cross-symbols. Their data has a similar 
	limiting magnitude and the data points can be compared directly. 
	The galaxy light profile in T1 (in arbitrarily scaled intensities,
	see Sect.\,5.2) is shown as dashed line in the upper panel. Power-law fits
	are shown as straight dashed lines. 
	Open circles mark the point source density profile uncorrected for 
	background. The backgrounds are indicated by the
	horizontal dotted lines.}
 \label{dirsch.fig8}
\end{figure}

To determine the radial density distribution we accounted for the radially variable
completeness by constructing the LF of all point sources brighter than T1=24 and
correcting the LF for completeness. The LF is then summed, normalized to the area
and background corrected. Beyond $\mathrm{r}>13\farcm5$ the number density of the
point sources in the whole/blue/red color interval remains constant and hence it has
been chosen as background. The used background densities are
$2.33\pm0.07/1.88\pm0.06/0.76\pm0.04 [\mathrm{obj}/\mathrm{arcmin}^2]$ for the
entire/blue/red sample. The resulting radial profiles are shown in
Fig.\,\ref{dirsch.fig8} and are tabulated in Table\,\ref{tab:radialGC}. In addition,
the values uncorrected for background and the background values are plotted in
Fig.\,\ref{dirsch.fig8} which illustrates the uncertainties of the employed
backgrounds.

In all radial distributions shown in Fig.\,\ref{dirsch.fig8} a change in slope at a
radial distance between 7\arcmin \ and 9\arcmin \ is apparent.  This steepening
occurs approximately two arc-minutes closer in for the red than for the blue
sub-population.  It is already visible for the blue population in the uncorrected
sample, however not as clearly. As it is customary we fit power-laws to the data
which are shown in Fig.\,\ref{dirsch.fig8} as dashed lines. To account for the
variable exponents that manifest themselves in the different slopes in the double
logarithmic presentation we split the radial range in smaller parts. The resulting
exponents ($\mathrm{n}(\mathrm{r})\propto r^\alpha$) are tabulated in
Table\,\ref{tab:radialfist}. While the blue clusters can be described with one
exponent within $1\arcmin-8\arcmin$ the density distribution of the red ones changes
the exponent within this radial interval.  The uncertainties of the power-law indices
are given for the background used (including the statistical error). However, a
systematic under- or overestimation of the background density might significantly
change the slopes for the larger radius interval. Since an underestimation is not
very probable considering Fig.\,\ref{dirsch.fig7}, the main uncertainty is an
overestimation, i.e. the cluster system is more extended. However, beyond 13\arcmin \
a radial dependence is not visible in any of the sub-samples nor does the color
distribution of the point sources show the characteristic peaks of the GCS.

\begin{table}
\begin{tabular}{cccc}
\hline
\hline
r[arcmin] & all clusters &  blue clusters &  red clusters \\\hline
$1\arcmin-8\arcmin$ & $-1.44\pm0.08$ & $-1.52\pm0.04$ & \\
$8\arcmin-12\arcmin$ & $-4.9\pm0.3$ & $ -5.1\pm0.3$ &\\
$1\arcmin-3\farcm5$ & & & $-0.78\pm0.04$ \\
$3\farcm5-6\arcmin$ & & & $-1.73\pm0.14$ \\
$6\arcmin-12\arcmin$ & & & $-4.9\pm0.4$ \\
\hline
\end{tabular}
\caption{This table lists the exponents of power-law fits
($\mathrm{n}(r)\propto r^\alpha$) to the cluster surface density
determined for different radial intervals.
The blue clusters can be described by two exponents, while three exponents
are needed for the red cluster and the complete sample.}
\label{tab:radialfist}
\end{table}

In the uppermost panel of Fig.\,\ref{dirsch.fig8} we also included the density
profile by Kissler et al. (\cite{kissler94}).  They used V measurements with a
limiting magnitude of 24. Since $\mathrm{V-R}\approx0.5$ (Galactic globular clusters
with C-T1=1.5)  this corresponds roughly to our brightness limit.  Except for the
last point the agreement is excellent. The last point deviates since they apparently
overestimated their background because they used a too small CCD field for a reliable
background determination. We also included in Fig.\,\ref{dirsch.fig8} the galaxy
light profile measured in the T1 filter (see the next section).

The observed density profile can be described with the projection of a truncated
power-law model with a core and a uniform exponent $\alpha$ (truncation radius
$\mathrm{r}_\mathrm{t}$, core radius $\mathrm{r}_\mathrm{c}$). Within the core a
constant number density is assumed. The resulting distributions are shown in
Fig.\,\ref{dirsch.fig8} for the red and the blue populations. In particular for the
blue population such a simple model provides an excellent fit to the observations.
The (used) parameters for the blue/red populations are: $\alpha=-2.3/-2$,
$\mathrm{r}_\mathrm{c}=1\farcm1/1\farcm5$ and
$\mathrm{r}_\mathrm{t}=12\arcmin/10\arcmin$.

Finally, one might ask whether the ''intermediate'' red sample
($1.55<(\mathrm{C-T1})<1.8$)  has the same distribution as the red and blue sample.  
Its radial distribution ($\alpha = -1.5\pm0.2$, $r<7\arcmin$)  is indistinguishable
within the uncertainties from these other distributions.

\subsection{Galaxy profile}

\begin{figure}[t]
 \centerline{\resizebox{\hsize}{!}{\includegraphics{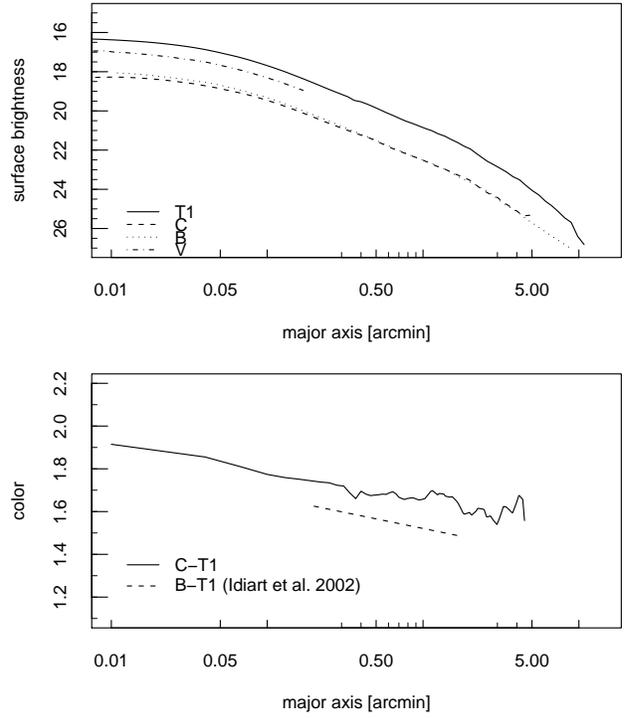}}}
 \caption{In the upper panel we show the galaxy luminosity profile
	measured in T1 (solid line) and C (dashed line). The dotted 
	line shows the B profile determined by Caon et al. (\cite{caon94})
	and the dash-dotted line the V profile from Lauer et al. 
	(\cite{lauer95}).
	In the lower panel the resulting color profiles are shown. 
	A solid line is used for C-T1 and a dotted one for B-T1.}
 \label{fig:gal_prof}
\end{figure}

The surroundings of NGC\,4636 are devoid of any large galaxy or bright foreground
stars and it is straightforward to use the {\it ellipse} task in IRAF/STSDAS to model
the galaxy. In Fig.\,\ref{fig:gal_prof} the luminosity and color profiles are shown
and tabulated in Table\,\ref{tab:galprof}. Caon et al. (\cite{caon94}) presented a B
luminosity profile for this galaxy which we include in our figure.  The V-T1 color
resulting from the surface profile of Lauer et al. (\cite{lauer95}) is
V-T1$=0.60\pm0.05$, which is in excellent agreement with the color given in Prugniel
et al.'s (\cite{prugniel98}) compilation ($0.58\pm0.02$). This agreement supports the
photometric calibration of T1.  The same holds within $\mathrm{r}<0\farcm5$ for the
B-T1 color that results from Caon et al.'s profile. However, for larger radii Caon et
al.'s profile is slightly shallower and the B-T1 color becomes bluer than C-T1.
However, this behavior is not supported by the color profile of Idiart et al.
(\cite{idiart02}).

A good empirical fit to the T1 profile is
$$
\mathrm{T1}(r) = -2.5\log\left(\mathrm{a}_1\left(1+\frac{r}{\mathrm{r}_{1}}\right)^{-\alpha_1}+
\mathrm{a}_2\left(1+\frac{r}{\mathrm{r}_{2}}\right)^{-\alpha_2}\right)
$$
with $\mathrm{a}_1=3.3\cdot 10^{-7}$, $\mathrm{a}_2=5.5\cdot 10^{-9}$,
$\mathrm{r}_{1}=0.11\arcmin$, $\mathrm{r}_{2}=8.5\arcmin$, $\alpha_1=2.2$, $\alpha_2=7.5$.
The standard deviation between the fit and the T1 profile is 0.037 and the maximal
deviations less than 0.09\,mag ($r<8\arcmin$).

\begin{figure}[t]
 \centerline{\resizebox{\hsize}{!}{\includegraphics{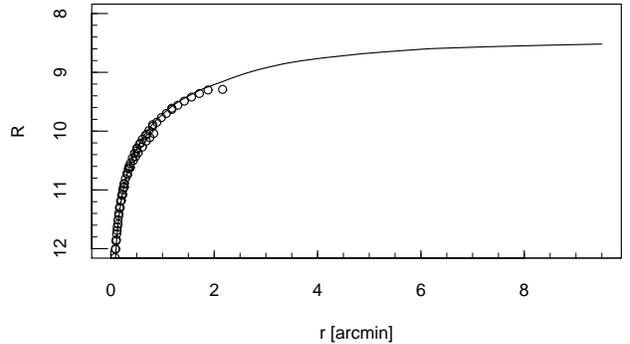}}}
 \caption{The integrated luminosity profile is compared to aperture measurements
	collected by Prugniel et al. (\cite{prugniel98}).} 
 \label{fig:prof_integ}
\end{figure}

To further check our photometric calibration we compare the integrated galaxy light
with aperture measurements compiled by Prugniel et al. (\cite{prugniel98}), shown in
Fig.\,\ref{fig:prof_integ}. For this comparison we transformed our T1 measurements
into R.  The overall agreement is very good. However, for radii larger than
$0\farcm5$, the measurements from the literature seem to be 0.03 magnitudes fainter
than ours. Since this shift is within our calibration uncertainty we make no attempt
to correct for it.

In Fig.\,\ref{dirsch.fig8} it is obvious that for radii smaller than 7\arcmin \ the
cluster population has a distinctly shallower distribution than the diffuse light, a
fact that has already been noted for the inner 4\arcmin \ by Kissler et al.
(\cite{kissler94}).  At larger radii, the slope of the galaxy light and cluster
density profile appear very similar; however, the uncertainties in the background
determination make a firm statement difficult. Beyond 11\arcmin \ (which corresponds
to $\approx 60$\,kpc) the stellar population of NGC\,4636 is negligible especially
compared to the X-ray halo which reaches 300\,kpc (Matsushita et al.
\cite{matsushita98}).

The color gradient of NGC\,4636 is relatively shallow (for a comparison with a galaxy
that has a strong radial color gradient see for example NGC\,1399, Dirsch et al.
\cite{dirsch03a}). Such a gradient can be caused by differences in the distribution
of distinct stellar populations, however, in NGC\,4636 high resolution archival WFPC2
HST images reveal an abundance of patchy dust structures in the inner region.
Therefore, reddening can play a major role in producing the color gradient. Hence,
the stellar populations seem to be well mixed.

\subsection{Azimuthal distribution}

To determine the ellipticity of the cluster system we study the azimuthal density
distribution of GC candidates in circular annuli.  The ellipticity $\epsilon$ and the
number density along the major and the minor axis ($\mathrm{N}_a, \mathrm{N}_b$) are
related via $\epsilon = 1-\left(\mathrm{N}_b/\mathrm{N}_a\right)^{1/\alpha}$, where
$\alpha$ is the exponent of the radial density distribution ($r^{-\alpha}$).

\begin{figure}[t]
 \centerline{\resizebox{\hsize}{!}{\includegraphics{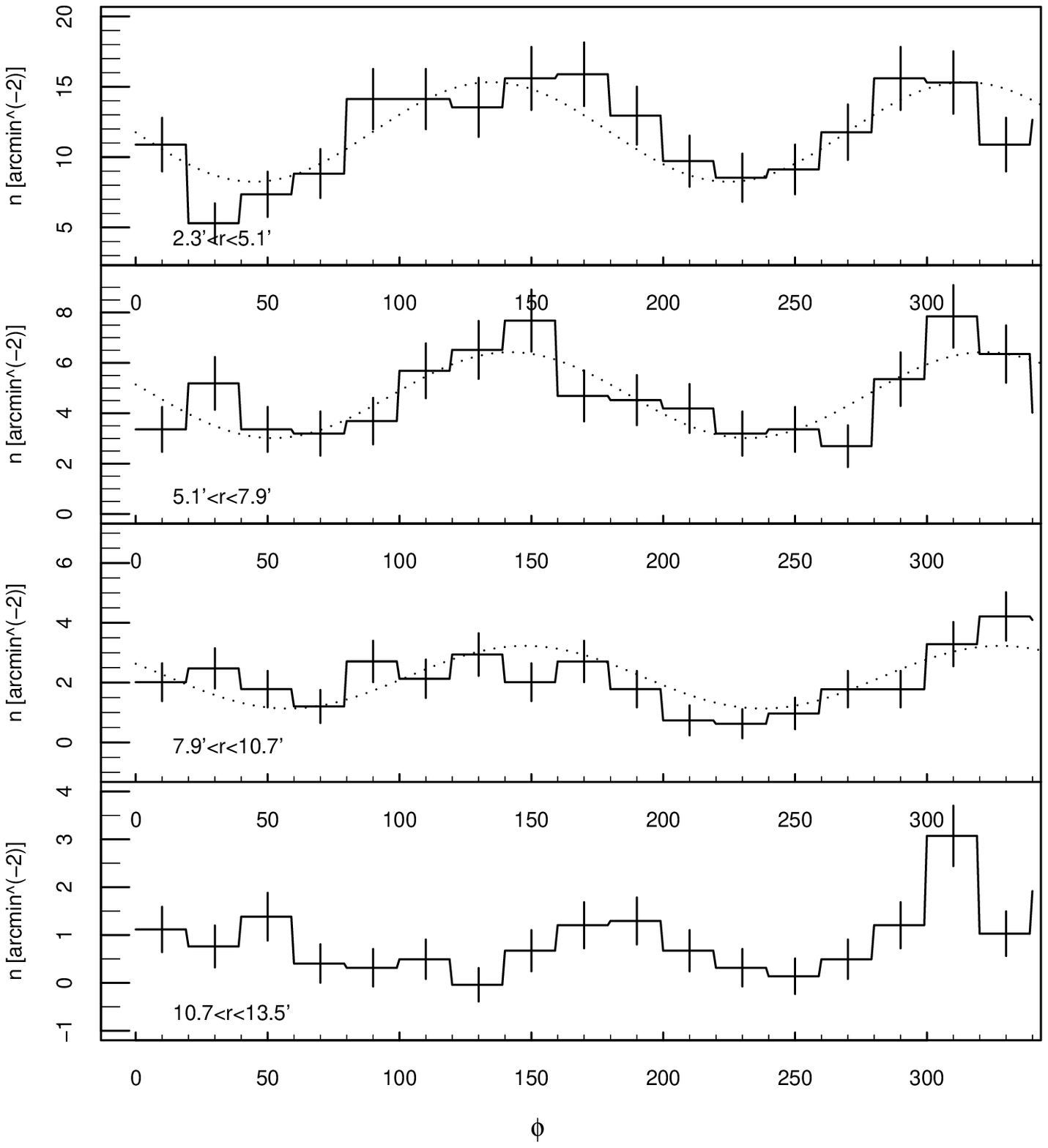}}}
 \caption{The azimuthal distribution of the cluster candidates 
	($0.9<(\mathrm{C-T1})<2.1$, $20<\mathrm{T1}<24$) is plotted for 
	four different radial bins ($2\farcm3-5\farcm1, 5\farcm1-7\farcm9, 
	7\farcm9-10\farcm7,10\farcm7-13\farcm5$). The dotted line shows 
	the fit of a function $\mathrm{a}\sin(2\phi+\mathrm{b})+\mathrm{c}$,
	which describes the number counts of an elliptically distributed 
	population that are counted in circular bins. The position angle is 
	counted from North to East.}
 \label{fig:azi_radial}
\end{figure}

In Fig.\,\ref{fig:azi_radial} the azimuthal cluster density is shown for four
different radial bins. In such a graph an elliptical cluster distribution manifests
itself as a $\sin(2\phi+\phi_0)$ distribution. We fit this function to the data which
is shown in Fig.\,\ref{fig:azi_radial} as the dotted line. Within $2\farcm3$ and
$7\farcm9$ the distribution is clearly elliptical. Within a radial range of
$7\farcm9$ and $10\farcm7$ a fit also results in an elliptical distribution, however,
the distribution is considerably asymmetric and the fit is relatively poor. At even
larger radii the distribution appears rather asymmetric with an enhancement of
clusters towards the North-West (around $310\degr$). This excess can be interpreted
as a slightly more extended cluster distribution towards this direction.
Interestingly, the galaxy light also shows an indication for such an asymmetric
elongation, however, flatfield uncertainties prohibit a definite answer, see
Fig.\,\ref{fig:n4636asym}.

\begin{figure}[t]
 \centerline{\resizebox{\hsize}{!}{\includegraphics{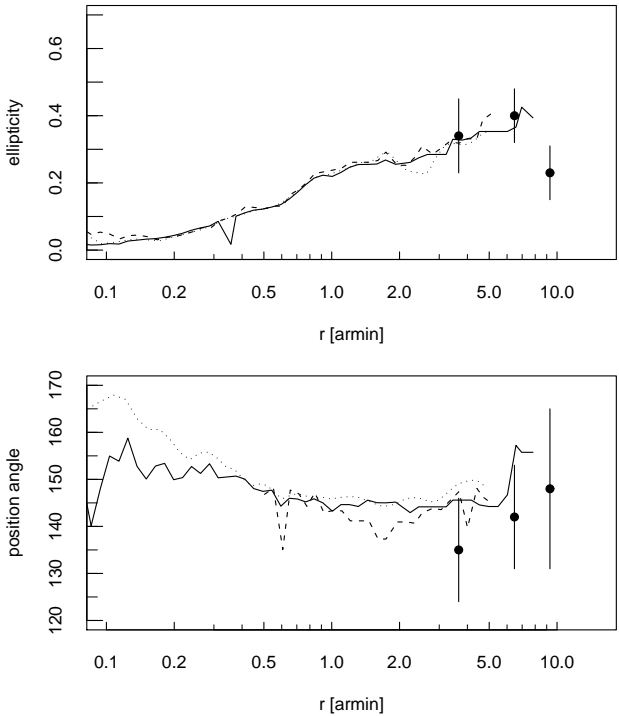}}}
 \caption{In this figure we compare in the {\bf upper panel} the ellipticity of
	the cluster system, measured in three radial bins, with the ellipticity 
	of the galaxy light determined in the R image (solid line),
	the C image (dashed line) and in a B image by Caon et al. (\cite{caon94}, dotted line).
	In the {\bf lower panel} the same is done for the 
	position angle. As x-axis we used the geometrical mean of semi-major and
	semi-minor axis.}
 \label{fig:azi_ellip_lumGC}
\end{figure}

\begin{figure}[t]
 \centerline{\resizebox{\hsize}{!}{\includegraphics[height=5cm]{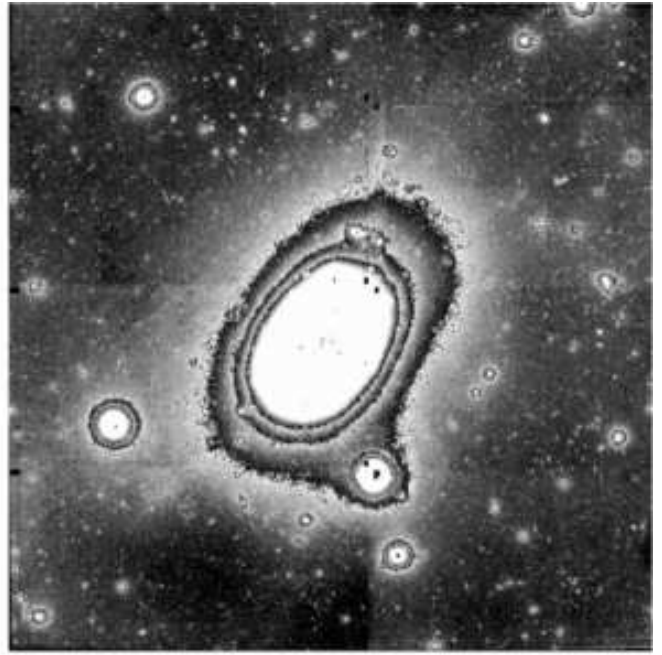}}}
 \caption{The R image of NGC\,4636 is shown ($33\arcmin\times33\arcmin$), North is up, East to 
	the left. The galaxy light appears to be more extended towards the North-West.}
 \label{fig:n4636asym}
\end{figure}

The radial dependence of the ellipticity and the position angle is shown with filled
circles in Fig.\ref{fig:azi_ellip_lumGC}. Within 8\arcmin \ the cluster system has an
ellipticity comparable to that of the galaxy light ($\epsilon\approx0.35$). At larger
radii no information for the galaxy light is available and the cluster ellipticity
drops to $0.23\pm0.09$, only a $1\sigma$ deviation from the ellipticity found at for
the smaller radii.

It is also of interest to study whether a difference in the azimuthal distribution of
red and blue clusters can be found. For this task we used all blue/red clusters
within $2\farcm3$ and $8\arcmin$, the radial range where the two samples are
indistinguishable. We found that the values for ellipticity and position angle agree
with the values obtained for the total sample. Also the ''intermediate'' color sample
defined above has the same azimuthal distribution within the uncertainties.

\section{The total number of clusters and the specific frequency}

To determine the total number of clusters we used the TOM and the width of the
Gaussian that had been fit to the GC luminosity function in Sect.\,4
($\mathrm{TOM}(\mathrm{T1})=23.31$). With this model we extrapolate from the number
of observed GCs to all clusters. We proceeded as follows: we summed all clusters
that are brighter than T1=24 in radial bins. These cluster counts were completeness
corrected to account for the radially changing completeness. In this way we observed
66\% of the total number of clusters. The final number of all clusters within a
radial distance of 14\arcmin \ is $4200\pm120$.

Kissler et al. (\cite{kissler94}) found $3600 \pm 500$ within $7\arcmin$, where we
find $3500 \pm 170$, which is in excellent agreement.

The specific frequency is the number of globular clusters per unit luminosity scaled
to a galaxy with an absolute luminosity of $\mathrm{M}_\mathrm{V}=-15$\,mag
($\mathrm{S}_\mathrm{N}=\mathrm{N}\cdot 10^{0.4(\mathrm{M}_\mathrm{V}+15)}$).  In
Sect.\,5.2 we have shown that $\mathrm{V}-\mathrm{T1}=0.60\pm0.05$.  The light
profile can be observed to a semi-major axis distance of $9\farcm5$. The GCs are
counted in circular rings and for the comparison we need the galaxy light in circular
rings as well. The largest aperture for which we can determine a reliable luminosity
is $7\farcm8$.  At this radius the total T1 luminosity of NGC\,4636 is
$\mathrm{T1}=8.70\pm0.05$. This is the maximum radius for which a specific frequency
can be obtained. The total number of cluster candidates within this radius is
$3570\pm180$. Hence we find $\mathrm{S}_\mathrm{N}=5.8\pm1.2$ and
$\mathrm{S}_\mathrm{N}=8.9\pm1.2$ for the GCS and the SBF distance, respectively.  A
slight inconsistency in these values is that we used the same number of clusters and
varied only the galaxy's luminosity. However, since the TOM is visible the number of
clusters does not change for the shorter distance.

\begin{figure}[t]
 \centerline{\resizebox{\hsize}{!}{\includegraphics{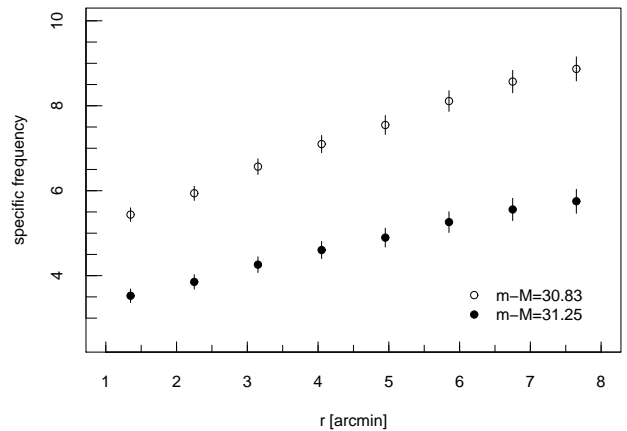}}}
 \caption{The specific frequency is plotted for different limiting radii. Open
	circles are used for a distance modulus of $30.83$
	(Tonry et al. \cite{tonry01}) and filled circles for 
	$(\mathrm{m}-\mathrm{M})=31.25$ (GCLF). The errors are only statistical errors.
	The distance uncertainties result in an additional uncertainty of 
	$\Delta\mathrm{S}_\mathrm{N}=1.2$.}
 \label{fig:sn}
\end{figure}

Since the clusters have a shallower distribution than the diffuse galaxy light the
specific frequency varies radially and it is not sufficient to give only one number
for the specific frequency.  Therefore we calculated the specific frequency for
different limiting radii for the distance of Tonry et al. (\cite{tonry01}) and ours.
The result is shown in Fig.\,\ref{fig:sn} and tabulated in Table\,\ref{tab:Sn}.

\begin{table}
\begin{tabular}{ccc}
\hline
\hline
 & $\mathrm{S}_\mathrm{N}$ & $\mathrm{S}_\mathrm{N}$ \\
r[arcmin] & $(\mathrm{m}-\mathrm{M})=31.25$ & $(\mathrm{m}-\mathrm{M})=30.83$ \\\hline
 0.5 &$ 3.7 \pm 0.2 $&$ 5.7 \pm 0.3 $ \\
 1.4 &$ 3.5 \pm 0.2 $&$ 5.4 \pm 0.3 $ \\
 2.3 &$ 3.9 \pm 0.2 $&$ 5.9 \pm 0.3 $ \\
 3.2 &$ 4.3 \pm 0.2 $&$ 6.5 \pm 0.3 $ \\
 4.1 &$ 4.6 \pm 0.2 $&$ 7.1 \pm 0.3 $ \\
 5.0 &$ 4.9 \pm 0.2 $&$ 7.5 \pm 0.3 $ \\
 5.9 &$ 5.3 \pm 0.3 $&$ 8.0 \pm 0.4 $ \\
 6.8 &$ 5.6 \pm 0.3 $&$ 8.5 \pm 0.4 $ \\
 7.7 &$ 5.8 \pm 0.3 $&$ 8.9 \pm 0.4 $ \\
\hline
\end{tabular}
\caption{Specific frequency based on a V-R=0.6 color without the
	the distance uncertainty that is constant 
	$\Delta\mathrm{S}_\mathrm{N}=1.2$.}
\label{tab:Sn}
\end{table}

The specific frequency is rather large, in particular if one uses the smaller
distance of Tonry et al. (\cite{tonry01}). In this case NGC\,4636 would have one of
the largest specific frequencies ever observed. In our opinion this is a further
argument against this shorter distance.

\section{Discussion}

The properties of the GCS of NGC 4636 present several puzzles. In the following
discussion we put them into a broader context. While definitive answers cannot be
given, we attempt to work out the environment inside which possible solutions can be
searched for. We summarize our results in Table\,\ref{tab:basicdata}.

\begin{table}
\begin{tabular}{ll}
\hline
\hline
Coordinates (J2000): & 
$12^\mathrm{h}42^\mathrm{m}49.9^\mathrm{s}$,  $2\degr41\arcmin16\arcsec$ \\
Assumed reddening: & $\mathrm{E}_\mathrm{C-T1} = 0.04\pm0.02$ \\
TOM distance modulus: & $(\mathrm{m}_\mathrm{M}) = 31.24\pm 0.17$\\
SBF distance modulus: & $(\mathrm{m}_\mathrm{M}) = 30.83 \pm 0.13$\\
Total number of GCs: & $\mathrm{N} = 4200 \pm 120$\\
Total apparent magnitude: & $\mathrm{T1} = 8.70\pm0.05$\\
Average color: & $\mathrm{V}-\mathrm{T1} = 0.60\pm0.05$\\
Global specific frequency: & \\
TOM distance:  & $\mathrm{S}_\mathrm{N} = 5.8\pm1.2 $\\
SBF distance:  & $\mathrm{S}_\mathrm{N} = 8.9\pm1.2 $\\
\hline
\end{tabular}
\caption{Basic data for NGC\,4636 used from the literature or derived in 
	this work}
\label{tab:basicdata}
\end{table}

\subsection{The specific frequency}

One of these puzzles is the specific frequency, which is surprisingly high for a
relatively isolated galaxy, even if our large distance modulus of 31.25 is correct. A
distance modulus of 30.8, as suggested by the SBF method, shifts
$\mathrm{S}_\mathrm{N}$ to unrivaled values. Richtler (\cite{richtler03})  compared
SBF distances and GCLF distances for a large sample of early-type galaxies. Apart
from an overall good agreement, deviations are almost exclusively found in the sense
of smaller SBF distances. These galaxies frequently exhibit other indicators pointing
to the existence of intermediate-age or even young stellar populations. The
difference of 0.4\,mag between SBF and GCLF moduli, as found for NGC 4636, is only
marginally explainable by the quoted measurement uncertainties. Moreover, the
appearance of a supernova (Giclas \cite{giclas39}) is another indication for an
intermediate-age population.  On the other hand, in those galaxies where younger
populations have unambiguously been identified, for instance in NGC\,1316 (Goudfrooij
et al. \cite{goudfrooij01})  or NGC\,5018 (Hilker\,\&\,Kissler-Patig \cite{hilker96},
Leonardi\,\&\,Worthey \cite{leonardi00}), the specific frequency is distinctly {\it
lower} due to the higher luminosity of younger populations (Gomez\,\&\,Richtler
\cite{gomez01}).  It is therefore difficult to explain the difference between SBF and
GCLF moduli by a contribution of an intermediate-age population to the SBF magnitude.
A small contribution would not be sufficient given the insensitivity of the I-band
fluctuation magnitude to population differences (e.g Cantiello et al.
\cite{cantiello03}), while a large one is improbable because of the high specific
frequency. The specific frequency of NGC 4636 is comparable to that of NGC 1399,
the central galaxy in the Fornax cluster (Dirsch et al. \cite{dirsch03a}). This
demonstrates that a central position in a high density environment is not a necessary
condition for developing a high $S_N$-value. Scenarios like stripping globular
clusters from neighboring galaxies, which have been constructed to explain the wealth
of clusters for example in NGC 1399, are not applicable in the case of NGC 4636.

The extent and brightness of the X-ray halo as well as the total mass indicated by
the X-ray analysis could also be characteristic of a small group of galaxies.
Together with the high specific frequency, one may ask whether NGC 4636 is the
remnant of a galaxy group. Isolated elliptical galaxies have several times been
suspected to result from multiple merging of small groups (e.g. Stocke et al.
\cite{stocke04}, Mulchaey \& Zabludoff \cite{mulchaey99}, Vikhlinin et al.
\cite{vikh99}). Numerical simulations of these merger processes show a strong
dependence on the initial conditions (Athanassoula et al. \cite{athana97}),
suggesting that the merger rate for group galaxies encompassed by a common dark
matter halo is low, the corresponding timescales possibly exceeding a Hubble time.
Thus NGC 4636 might be the merger outcome of a close group of galaxies with
individual dark halos and the present halo the dispersed debris of the merger
components.  More properties, which support such a scenario, are the unusually high
dust content (Temi et al. \cite{temi03}) and the absence of a color gradient which
suggests that chemical evolution was largely terminated at the formation time of NGC
4636.

The high, perhaps extremely high, star formation rate that accompanied the merger
is favorable for a high efficiency of cluster formation (Larsen\,\&\,Richtler
\cite{larsen00}), giving a qualitative explanation for the high specific frequency of
NGC 4636. Searching for possible counterparts, one comes across NGC\,1132, which has
been suspected of being the merger remnant of a small galaxy group (Mulchaey \&
Zabludoff \cite{mulchaey99}). However, nothing is known about its cluster system due
to its large distance of about 100 Mpc.

McLaughlin (\cite{mclaughlin99}) argued that the efficiency of GC formation is
constant if one normalizes the total cluster mass to the sum of stellar mass and
X-ray gas mass. Does the high specific frequency come back to ''normal'' when the gas
mass contained in the X-ray halo of NGC 4636 is included? For this, one needs factors
of the order 1.5-2. One problem is the uncertain volume, inside which this
normalization has to take place. The GCS terminates at a radius of roughly 40 kpc.
According to Matsushita et al. (\cite{matsushita98}), the stellar mass inside this
radius still is a factor of 10 larger than the X-ray gas mass, rendering the
inclusion of the gas mass insignificant. Beyond this radius, the galaxy's light
cannot be traced any more.  Only at a radius of 200 kpc, apparently much larger than
the extension of the GCS and the galaxy, the X-ray gas mass becomes comparable to the
total stellar mass. The question remains how the history of GC formation and the
X-ray gas are related. A more or less constant ratio of GC mass and
total mass does not contradict a {\it local} high efficiency of GC formation in a
merger event.

\subsection{The luminosity function}

It has been theoretically predicted by Ashman et al. (\cite{ashman95}) that, for a
given break in the mass function, metal-poor clusters have a brighter TOM than more
metal-rich clusters.  Such differences have also been observed in the V luminosity 
function of several
GCS by Larsen et al. (\cite{larsen01}). Despite those expectations we do not see
convincing evidence for this effect in NGC\,4636. For example, considering the blue
and the very red population (see Table\,\ref{tab:lktfits}) and using the Washington
color-metallicity relation of Harris\,\&\,Harris (\cite{harris02}), we find a mean
metallicity of -1.44\,dex and -0.08\,dex for these populations. Such a metallicity
difference should manifest itself in a TOM difference of $\Delta{T1} \approx 0.3$ (we
used the fact that T1 is basically R and Table\,3 of Ashman et al. \cite{ashman95}).
Such a difference cannot be seen in Table\,\ref{tab:lktfits}. However, the fitted
width of the Gaussian is much smaller for the very red than for the blue sample. If
we had used the same width we would have found a TOM for the whole radial range of
$\mathrm{T1} = 23.46\pm0.18$ which is also within the uncertainties equal to that of
the blue sample. We have no good explanation for the absence of the predicted TOM
dependence on metallicity.

We have shown that the intermediate color cluster sample has a LF that differs from
that of the other sub-populations in the sense that it has brighter GCs
(Fig.\,\ref{dirsch.fig6}). This behavior can also be seen in the brightness selected
color distributions (Fig.\,\ref{dirsch.fig4}). It is interesting to note that this is
similar to the GCS around NGC\,1399, where the brighter GCs have a different color
distribution (Dirsch et al. \cite{dirsch03a}, Ostrov et al. \cite{ostrov98}). In
particular the bright GCs have no bimodal color distribution any longer but peak
at an intermediate color. We tentatively interpret both observations with a picture
in which the ``normal'' GCs are mixed with a population of objects that are brighter
and of intermediate color. These may be former nuclei
of stripped nucleated dwarf galaxies.

\subsection{The ''edge'' of NGC\,4636}

The GC density profile becomes very steep beyond $8\arcmin$ (see
Fig.\,\ref{dirsch.fig8}). Since the GCS traces the galaxy light (see also the
Section\,7.4), it is reasonable to assume that the whole stellar body of NGC\,4636
shows this behavior, which is supported by a relatively sharp border of the galaxy
light. Such steepening in the very outer regions of elliptical galaxies has been
explained by Jaffe (\cite{jaffe87}) and White (\cite{white87}) on the basis of energy
considerations. They argue that the transition from an $r^{-3}$ to an $r^{-4}$
spatial density profile occurs if there is a sharp cut in the energy distribution
function near the escape energy in the case of an approximately Keplerian potential.

A different case is, for example, NGC\,1399, which has also been studied with
wide-field CCD photometry by Dirsch et al. (\cite{dirsch03a}). They attributed its
(spatially) uniform $r^{-3}$ profile to its extended dark matter halo. The Keplerian
regime (if it exists at all in the inner region of the Fornax cluster) would then be
beyond the observable radial range. If this idea was applicable to NGC\,4636 it would
imply that we have already reached the regime of a declining potential.
 
However, this interpretation conflicts with the existence of a massive dark halo - as
is claimed on the basis of X-ray observations (Loewenstein\,\&\,Mushotzky
\cite{loewenstein03}, Matsushita et al. \cite{matsushita98}, Mushotzky et al.
\cite{mushotzky94}). Hence either our interpretation of the light profile and of the
cluster distribution is incorrect or the X-ray mass profile is inaccurate. Problems concerning the
interpretation of the X-ray gas arise from its turbulent nature (Otho et al.
\cite{otho03}, Jones et al. \cite{jones02}) and a possible interaction with the Virgo
inter-cluster gas (Trinchieri et al. \cite{trinchieri94}). Therefore an independent
measurement of the dark matter distribution is desirable.

\subsection{Is the GC radial density distribution peculiar?}

The radial distribution of the GCS in NGC 4636 is considerably shallower than the
galaxy light over nearly the entire radial range. How can this be assessed in a
quantitative manner? GCSs generally exhibit shallower
surface density profiles than their host galaxies' light (Harris \cite{harris01},
Ashman\,\&\,Zepf \cite{ashman98}). It is also known that larger galaxies tend to have
GCSs with shallower radial distributions (Kissler-Patig \cite{kisslerpatig97},
Ashman\,\&\,Zepf \cite{ashman98}), which also holds for the galaxy profile
(Kormendy\,\&\,Djorgovski \cite{kormendy89}).  However, we are not aware of any
recent quantitative comparison examining these trends. In the following we collect
power-law indices fitted to the radial distribution of GCSs from the literature and
perform a similar, homogeneous fit to published galaxy light profiles within a
comparable radial range. The correlation of these power-law indices with galaxy
luminosity is determined and compared for GCSs and field stars.

\begin{figure}[t]
 \centerline{\resizebox{\hsize}{!}{\includegraphics{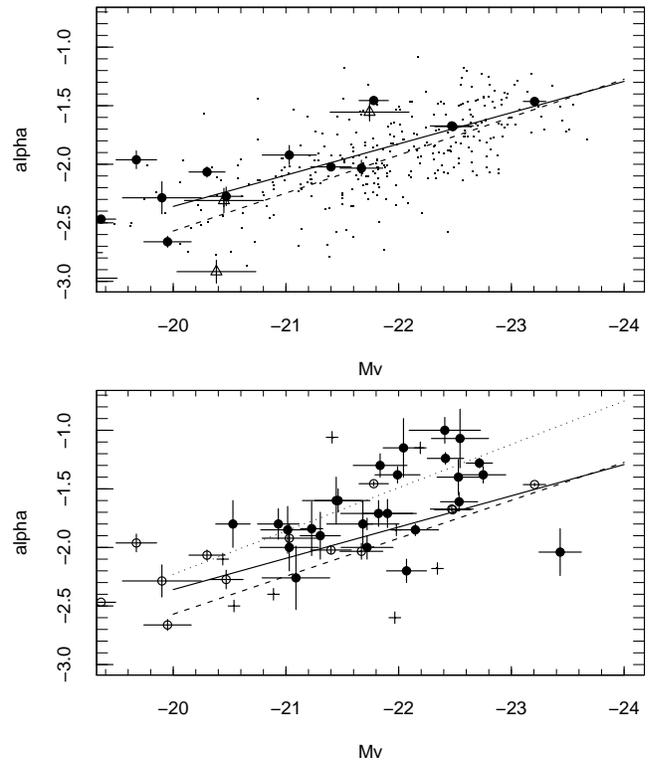}}}
 \caption{{\bf Upper panel:} the dependence of the power-law index of the
	surface brightness profile on galaxy luminosity (multiplied by
	-0.4) of Virgo/Fornax galaxies published by Caon et al. (\cite{caon94}) (see text for a
	detailed description). For the galaxies plotted as filled circles
	distance determinations via SBF exist (Tonry et al. \cite{tonry01})	
	while mean cluster distances have been assumed for the galaxies
	plotted with open triangles. The solid line is a fit to these 
	galaxies brighter than $\mathrm{M}_\mathrm{V}=-20$. The dots are similar 
	measurements obtained by Schombert et al. (\cite{schombert86}), 
	transformed to the same magnitude scale (see text). The dashed line
	shows the fit to these data (also brighter than $\mathrm{M}_\mathrm{V}=-20$).
	{\bf Lower panel:} the exponents of power-law fits to the GC density
	distributions of elliptical galaxies is plotted (filled circles)
	together with the galaxies from the upper sample having
	SBF distances (open circles).
	The crosses show the data of Kissler-Patig's compilation with 
	errors in the exponent larger than 0.3 or in which the distribution
	has been obtained with less than 4 data points.
	It is obvious that the GCs in general have
	a shallower distribution than the galaxy light and that the correlation 
	with the total galaxy luminosity is very similar between them.
	The data plotted with the filled symbols is tabulated in Table\,\ref{tab:mv_alpha1} and 
	Table\,\ref{tab:mv_alpha2}. The solid and dashed lines show the fits 
	to the galaxy light and correspond to the lines in the upper panel. The 
	dotted line shows a fit to the GCS radial distribution.}
 \label{dirsch.fig15}
\end{figure}

We use the galaxy profile measurements of Virgo and Fornax ellipticals published by
Caon et al. (\cite{caon94}) and fit power-laws to the data (we multiplied the
luminosity profiles by -0.4 to ensure comparability with the cluster data).  A
power-law fit may not be the preferred fit to a galaxy light profile. It is chosen
here to allow a comparison with the published cluster density profiles which are
generally described by power-laws. In order to exclude galaxies for which a single
power-law at radii larger than 20\arcsec \ is not appropriate, we fitted two radial
ranges and omitted those galaxies for which the power-law indices deviate by more
than 0.15. The B luminosity profiles from Caon et al. (\cite{caon94}) were
transformed to V magnitudes with B-V colors from Prugniel et al. (\cite{prugniel98}).
We adopted the total apparent magnitudes given by Caon et al.. For the absolute
magnitudes we used the distance moduli of Tonry et al. (\cite{tonry01}). For galaxies
without a SBF distance we used a mean cluster distance modulus of 31.4 for Fornax and
31.1 for Virgo. To correct for absorption the mean of the Schlegel et al.
(\cite{schlegel98}) and Burstein\,\&\,Heiles (\cite{burstein82}) values was employed.
The result is shown in the upper panel of Fig.\,\ref{dirsch.fig15}, which illustrates
the well-known fact that larger early-type galaxies are less concentrated (e.g.
Kormendy\,\&\,Djorgovski \cite{kormendy89}). In this plot we also included for a
consistency check the same dependence published by Schombert et al.
(\cite{schombert86}). This data, however, is much more heterogeneous and not
reddening corrected. The distances are those provided by Schombert et al.
(\cite{schombert86}) and are based on the recession velocities. Furthermore, the
galaxy luminosities published by Schombert et al. are measured within a linear radius
of 16\,kpc. To transform these luminosities to the scale of Caon et al. we compared
galaxies that are contained in both samples and obtained a mean shift of
$\mathrm{m}_{\mathrm{V\,Caon}} = \mathrm{m}_{\mathrm{V\,Schombert}} - 0.45$. One
expects a systematic trend of this shift with magnitude, however, we were unable to
detect it, presumably due to the small magnitude coverage of the galaxies in common
(12 objects). We fit a linear function to the galaxies brighter than
$\mathrm{M}_\mathrm{V}=-20$ of the two samples and found a slope of $-0.31\pm0.03$,
consistent with Schombert's sample. Considering its luminosity ($M_V
=-21.78\pm0.13$), NGC\,4636 has a surprisingly shallow light distribution
($\alpha=-1.63\pm0.05$).

To compare the dependence of the GCS radial distribution and the luminosity profile
on the total galaxy luminosity we compiled data from the literature, the most
important single source being Kissler-Patig (\cite{kisslerpatig97}) in conjunction
with distance moduli from Tonry et al. (\cite{tonry01}).  Furthermore, we excluded
those cases in which the error of the power-law exponent is larger than 0.3 and cases
in which the power-law fit is done on the basis of less than 4 points and included a
few galaxies with recent measurements. For the galaxy luminosity we have chosen the
value given by Caon et al. (\cite{caon94}) or, when not available, the total V
magnitude from the RC3. The resulting dependence of the cluster radial distribution
on the host galaxy's luminosity is shown in the lower panel of
Fig.\,\ref{dirsch.fig15}.

The comparison of the radial cluster distribution with the galaxy light shows the
well established fact that in general the cluster systems have shallower
distributions. However, it also shows that the difference between the slopes of
galaxy light and the GC radial profiles depends on luminosity: brighter galaxies
exhibit a slightly larger difference on the average, which was not yet obvious in the
sample discussed by Ashman \& Zepf \cite{ashman98}.

The radial distribution of the GCs of one elliptical is rather different: NGC\,1316,
the brightest galaxy in the sample, clearly stands out. However, it is not a typical
elliptical galaxy, but a merger remnant with an age of 2-3\,Gyr (Goudfrooij et al.
\cite{goudfrooij01}, Gomez\,\&\,Richtler \cite{gomez01}).

The location of NGC\,4636 in this diagram on the other hand is typical. 

In Table\,\ref{tab:mv_alpha2} we also included the difference between the slope of
the GC radial density profile and the galaxy light for several galaxies where both
numbers are available. It shows that both slopes are closely related, irrespective of
luminosity.

It is important to note that a {\it shallower density profile of the GCS does not
necessarily imply a GCS that is more extended than the galaxy light}. Both in
NGC\,4636 and NGC\,1399 the GCS has a clearly shallower profile than the galaxy light
until a certain radius. At large radii, however, the distribution becomes steeper and
is compatible with that of the galaxy light. The radius at which this happens is
fairly large and hence the shallower inner decline does not correspond to a core.

\subsection{The discrepancy between the distribution of clusters and field stars}

Another fact worth pointing out is that the ellipticities of the GCS and the field
population of NGC\,4636 are the same (within the uncertainties) within 4\arcmin \ and
9\arcmin \ while the radial distributions are rather different in this range.

Also puzzling is the fact that no pronounced radial color gradient can be seen for
the galaxy light. One might expect that any field component that is connected to the
metal-poor, blue cluster population (and which should therefore follow the same
density profile) would become radially more prominent and hence the galaxy light is
expected to have a stronger color gradient than actually observed.  The implication
is that the population that has formed together with the clusters (in particular the
metal-poor, blue ones) is negligible compared to the field population that formed
with a very low cluster formation efficiency. This is in strong contrast to other
galaxies where the clusters have a shallower distribution and in which strong color
gradients in the light profile have been found:  NGC\,1427 (Forte et al.
\cite{forte01}), NGC\,4472 (Rhode \& Zepf \cite{rhode01}), NGC\,1399 (Dirsch et al.
\cite{dirsch03a}).

\begin{figure}[t]
 \centerline{\resizebox{\hsize}{!}{\includegraphics{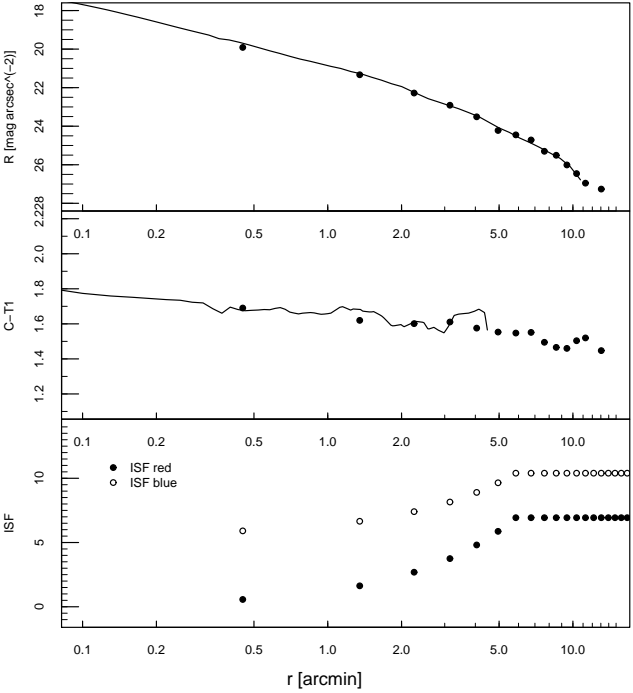}}}
 \caption{{\bf Upper-most panel:} the observed light profile (solid line) is 
	compared to the fitted light profile (solid circles), based on an assumed ISF, which is
	different for the red and blue clusters (see text for details).
	{\bf Middle panel:} the color profile resulting from the fit (solid circles) is
	shown together with the observed profile. {\bf Lower panel:} the ISF of the
	blue and the red population used to generate the luminosity and color profile is
	shown. The ISF values are transformed into the V system using a constant V-R color
	of 0.6. This has been done for an easier comparison with the values published in Dirsch
	et al. (\cite{dirsch03a}).}
 \label{dirsch.fig16}
\end{figure}

In the following we perform a study similar to that presented in Dirsch et al.
(\cite{dirsch03a}):  we aim at constraining the so-called {\it intrinsic specific
frequency} (ISF) which measures the specific frequencies of the red clusters with
respect to a red population of the same mean color and of the metal-poor clusters
with respect to their underlying field population. Such an approach results in a
unique solution only when the two populations are assumed and the two observables,
light profile and color profile, are fitted by varying only the blue and red ISF. The
result is shown in Fig.\,\ref{dirsch.fig16}. For radii larger than 6\arcmin \ a
constant ISF for each population can be used, which corresponds to the fact that
cluster and light distribution become similar at large radii. The ISF of the blue
population is, as expected, at all radii larger than that of the red ones.
Furthermore, at large radii the value of the blue population's ISF is very similar to
that derived for NGC\,1399. However, while in the latter galaxy the blue ISF remains
constant with radius, it declines in NGC\,4636 inwards. This decline is accompanied
by a decline in the red cluster density profile, otherwise a color gradient would
arise. The value of the red ISF is considerably smaller in NGC\,4636 than in
NGC\,1399.

Summarizing, we find that in particular the red clusters have formed with a lower
efficiency in NGC\,4636 compared to NGC\,1399, while it is slightly higher for the
blue clusters for radii greater than 6\arcmin. Within this radius also their
formation efficiency was lower. Despite these differences the global specific
frequencies are very similar and hence the overall cluster formation efficiencies
have been comparable.

\subsection{The globular cluster - X-ray gas connection}

The GCS and the hot X-ray gas may have a common evolutionary history (via mass loss
during star formation and SN explosions of the respective stellar population),
however, there is no reason to expect a tight relationship in their distribution. It
is therefore striking to find that similar radii appear to be characteristic of 
both. We have shown than the GC density distribution becomes steeper at
$6\arcmin-8\arcmin$ (depending on the population) which is also where the X-ray gas
changes its slope and becomes {\it flatter} (Trinchieri et al. \cite{trinchieri94}).
At a radius of about $13\arcmin$ the cluster system appears to have ended; this is
the radius where the intensity profile of the X-ray gas changes again and becomes
nearly flat (at $\approx 12$\,kpc, Matsushita et al. \cite{matsushita98}). Is this
merely a coincidence? A study of the dynamical properties of the GCS that can
constrain the matter distribution is required to shed light on these findings.

\appendix

\section{The data tables}

The luminosity function of the GCs, their color and radial distribution and the
galaxy profile including the ellipticity are published electronically. The
galaxy sample used in Sec.\,7.4 is also available electronically. The table captions are
given in this appendix.

\begin{table}[h]
\caption{
The position, the T1 magnitude and C-T1 color for each of the
point sources identified in the investigated frame.
}
\label{tab:data}
\end{table}

\begin{table}
\begin{tabular}{ccc}
\hline
\hline
C-T1 & N($3\farcm6<r<8\farcm1$) & N($8\farcm1<r<13\farcm5$) \\\hline
0.89 & $  4.2 \pm  5.8 $ & $ -7.9 \pm  9.7 $\\
0.96 & $  9.7 \pm  6.5 $ & $ 10.5 \pm 11.0 $\\
1.02 & $  7.0 \pm  7.4 $ & $ -3.1 \pm  9.7 $\\
1.10 & $ 23.4 \pm  7.7 $ & $ 29.2 \pm 12.4 $\\
1.17 & $ 63.8 \pm 10.0 $ & $ 31.1 \pm 12.4 $\\
1.24 & $ 83.6 \pm 11.0 $ & $ 35.4 \pm 12.8 $\\
1.31 & $ 75.8 \pm 10.5 $ & $ 49.1 \pm 13.1 $\\
1.38 & $ 79.0 \pm 10.5 $ & $ 46.3 \pm 12.3 $\\
1.45 & $ 61.7 \pm  9.4 $ & $ 17.5 \pm 10.4 $\\
1.52 & $ 39.6 \pm  7.7 $ & $ 16.5 \pm  9.3 $\\
1.58 & $ 40.8 \pm  7.9 $ & $  9.7 \pm  9.2 $\\
1.66 & $ 61.5 \pm  9.0 $ & $ 33.4 \pm 10.0 $\\
1.73 & $ 73.2 \pm  9.7 $ & $ -0.4 \pm  8.5 $\\
1.80 & $ 63.7 \pm  8.7 $ & $ 15.1 \pm  7.6 $\\
1.87 & $ 56.2 \pm  8.4 $ & $ 13.6 \pm  8.1 $\\
1.94 & $ 32.6 \pm  6.4 $ & $ 16.5 \pm  6.8 $\\
2.00 & $ 36.9 \pm  6.9 $ & $  9.8 \pm  6.8 $\\
2.08 & $ 11.3 \pm  4.6 $ & $  6.8 \pm  6.5 $\\
2.15 & $  7.3 \pm  4.2 $ & $ -1.2 \pm  5.8 $\\
2.21 & $ 10.0 \pm  4.3 $ & $ -0.6 \pm  5.3 $\\
\hline
\end{tabular}
\caption{The color distribution of the GC candidates brighter than
        R=24 is shown in
        Fig.\,\ref{dirsch.fig4}. In this table we used the
        histogram counts with a bin-width of 0.07\,mag.}
\label{tab:colordistr}
\end{table}

\begin{table*}
\begin{tabular}{ccccccc}
\hline
\hline
r & \multicolumn{2}{c}{n(all)[arcmin$^{-2}$]} &
\multicolumn{2}{c}{n(blue)[arcmin$^{-2}$]} &
\multicolumn{2}{c}{n(red)[arcmin$^{-2}$]} \\
 & with background & background subtracted & with background & background subtracted & with background & background subtracted \\
\hline
$ 0\arcmin45 $&$ 64.25\pm 3.39 $&$ 61.93\pm 3.37 $&$ 21.63\pm 2.15 $&$ 20.87\pm 2.15 $&$ 50.04\pm 2.92 $&$ 48.16\pm 2.92 $\\
$ 1\arcmin35 $&$ 38.59\pm 1.47 $&$ 36.26\pm 1.47 $&$ 14.43\pm 0.87 $&$ 13.67\pm 0.88 $&$ 26.81\pm 1.25 $&$ 24.94\pm 1.26 $\\
$ 2\arcmin25 $&$ 23.20\pm 0.69 $&$ 20.87\pm 0.69 $&$  9.71\pm 0.47 $&$  8.95\pm 0.48 $&$ 14.77\pm 0.54 $&$ 12.90\pm 0.55 $\\
$ 3\arcmin15 $&$ 16.70\pm 0.48 $&$ 14.36\pm 0.49 $&$  7.91\pm 0.33 $&$  7.16\pm 0.34 $&$  9.40\pm 0.37 $&$  7.52\pm 0.38 $\\
$ 4\arcmin05 $&$ 12.26\pm 0.54 $&$  9.93\pm 0.55 $&$  5.52\pm 0.35 $&$  4.76\pm 0.35 $&$  7.56\pm 0.47 $&$  5.69\pm 0.47 $\\
$ 4\arcmin95 $&$  8.90\pm 0.40 $&$  6.57\pm 0.40 $&$  3.56\pm 0.28 $&$  2.80\pm 0.28 $&$  5.44\pm 0.29 $&$  3.57\pm 0.29 $\\
$ 5\arcmin85 $&$  8.01\pm 0.41 $&$  5.69\pm 0.41 $&$  3.39\pm 0.26 $&$  2.63\pm 0.26 $&$  5.07\pm 0.34 $&$  3.20\pm 0.35 $\\
$ 6\arcmin75 $&$  6.72\pm 0.33 $&$  4.39\pm 0.34 $&$  2.85\pm 0.22 $&$  2.09\pm 0.22 $&$  4.34\pm 0.28 $&$  2.46\pm 0.28 $\\
$ 7\arcmin65 $&$  5.01\pm 0.25 $&$  2.68\pm 0.26 $&$  1.72\pm 0.15 $&$  0.97\pm 0.15 $&$  3.68\pm 0.22 $&$  1.81\pm 0.23 $\\
$ 8\arcmin55 $&$  4.68\pm 0.23 $&$  2.35\pm 0.24 $&$  1.45\pm 0.13 $&$  0.69\pm 0.14 $&$  3.54\pm 0.20 $&$  1.66\pm 0.21 $\\
$ 9\arcmin45 $&$  3.84\pm 0.20 $&$  1.51\pm 0.21 $&$  1.18\pm 0.12 $&$  0.42\pm 0.12 $&$  2.94\pm 0.18 $&$  1.07\pm 0.19 $\\
$10\arcmin35 $&$  3.30\pm 0.21 $&$  0.98\pm 0.22 $&$  1.10\pm 0.12 $&$  0.35\pm 0.13 $&$  2.48\pm 0.18 $&$  0.60\pm 0.19 $\\
$11\arcmin25 $&$  3.00\pm 0.20 $&$  0.65\pm 0.19 $&$  0.99\pm 0.09 $&$  0.24\pm 0.10 $&$  2.23\pm 0.16 $&$  0.36\pm 0.17 $\\
$12\arcmin15 $&$  2.65\pm 0.17 $&$  0.32\pm 0.18 $&$  0.75\pm 0.09 $&$  0.00\pm 0.10 $&$  2.22\pm 0.15 $&$  0.34\pm 0.16 $\\
$13\arcmin05 $&$  2.85\pm 0.18 $&$  0.52\pm 0.20 $&$  0.93\pm 0.11 $&$  0.17\pm 0.12 $&$  2.35\pm 0.16 $&$  0.47\pm 0.18 $\\
$13\arcmin95 $&$  2.05\pm 0.15 $&$ -0.27\pm 0.16 $&$  0.70\pm 0.10 $&$ -0.05\pm 0.11 $&$  1.62\pm 0.13 $&$ -0.26\pm 0.14 $\\
$14\arcmin85 $&$  2.33\pm 0.16 $&$  0.01\pm 0.17 $&$  0.75\pm 0.09 $&$  0.00\pm 0.10 $&$  1.88\pm 0.15 $&$  0.00\pm 0.16 $\\
$15\arcmin75 $&$  3.15\pm 0.25 $&$  0.82\pm 0.26 $&$  0.85\pm 0.14 $&$  0.09\pm 0.15 $&$  2.77\pm 0.22 $&$  0.89\pm 0.23 $\\
$16\arcmin65 $&$  2.79\pm 0.23 $&$  0.46\pm 0.24 $&$  0.94\pm 0.13 $&$  0.18\pm 0.13 $&$  2.05\pm 0.20 $&$  0.17\pm 0.21 $\\
$17\arcmin55 $&$  3.19\pm 0.26 $&$  0.86\pm 0.27 $&$  1.07\pm 0.17 $&$  0.32\pm 0.17 $&$  2.47\pm 0.21 $&$  0.60\pm 0.22 $\\
$18\arcmin45 $&$  2.78\pm 0.31 $&$  0.45\pm 0.32 $&$  1.18\pm 0.22 $&$  0.42\pm 0.23 $&$  1.87\pm 0.25 $&$  0.00\pm 0.26 $\\
$19\arcmin35 $&$  2.02\pm 0.27 $&$ -0.31\pm 0.28 $&$  0.36\pm 0.14 $&$ -0.40\pm 0.14 $&$  2.12\pm 0.30 $&$  0.24\pm 0.30 $\\
\hline
\end{tabular}
\caption{The radial densities of all GCs and the red and blue subsamples
        with and without background subtraction
        as shown in Fig.\,\ref{dirsch.fig8}. The densities given here are
        calculated for clusters brighter T1=24. To include the entire
        luminosity function these values have to be multiplied by 1.52 (
        see Sect.\,5).}
\label{tab:radialGC}
\end{table*}

\begin{table}
\begin{tabular}{ccccc}
\hline
\hline
r  & T1 & C-T1 &  PA & $\epsilon$\\
& $[\mathrm{mag/arcsec}^{-2}]$ & & &\\\hline
$0\farcm00$ &$ 16.27 \pm 0.03 $&$ 1.91 \pm 0.05 $&$ 88 \pm 9 $&$ 0.31 \pm 0.08$\\
$0\farcm01$ &$ 16.34 \pm 0.03 $&$ 1.91 \pm 0.05 $&$144 \pm12 $&$ 0.04 \pm 0.02$\\
$0\farcm02$ &$ 16.46 \pm 0.03 $&$ 1.88 \pm 0.05 $&$108 \pm 9 $&$ 0.03 \pm 0.01$\\
$0\farcm03$ &$ 16.76 \pm 0.03 $&$ 1.86 \pm 0.05 $&$124 \pm 5 $&$ 0.03 \pm 0.01$\\
$0\farcm05$ &$ 17.01 \pm 0.03 $&$ 1.85 \pm 0.05 $&$120 \pm20 $&$ 0.00 \pm 0.01$\\
$0\farcm07$ &$ 17.33 \pm 0.03 $&$ 1.81 \pm 0.05 $&$157 \pm 3 $&$ 0.02 \pm 0.01$\\
$0\farcm10$ &$ 17.63 \pm 0.03 $&$ 1.77 \pm 0.05 $&$155 \pm 3 $&$ 0.02 \pm 0.01$\\
$0\farcm14$ &$ 17.97 \pm 0.03 $&$ 1.75 \pm 0.05 $&$153 \pm 2 $&$ 0.03 \pm 0.01$\\
$0\farcm18$ &$ 18.35 \pm 0.03 $&$ 1.74 \pm 0.05 $&$153 \pm 2 $&$ 0.04 \pm 0.01$\\
$0\farcm20$ &$ 18.61 \pm 0.03 $&$ 1.74 \pm 0.05 $&$150 \pm 2 $&$ 0.04 \pm 0.01$\\
$0\farcm26$ &$ 18.89 \pm 0.03 $&$ 1.73 \pm 0.05 $&$151 \pm 2 $&$ 0.06 \pm 0.01$\\
$0\farcm34$ &$ 19.27 \pm 0.03 $&$ 1.69 \pm 0.05 $&$150 \pm 2 $&$ 0.08 \pm 0.01$\\
$0\farcm45$ &$ 19.65 \pm 0.03 $&$ 1.67 \pm 0.05 $&$149 \pm 2 $&$ 0.11 \pm 0.01$\\
$0\farcm54$ &$ 19.94 \pm 0.03 $&$ 1.67 \pm 0.05 $&$147 \pm 2 $&$ 0.13 \pm 0.01$\\
$0\farcm65$ &$ 20.23 \pm 0.03 $&$ 1.67 \pm 0.05 $&$145 \pm 2 $&$ 0.14 \pm 0.01$\\
$0\farcm77$ &$ 20.52 \pm 0.03 $&$ 1.66 \pm 0.05 $&$146 \pm 2 $&$ 0.17 \pm 0.01$\\
$1\farcm04$ &$ 20.89 \pm 0.03 $&$ 1.66 \pm 0.05 $&$145 \pm 2 $&$ 0.22 \pm 0.01$\\
$1\farcm38$ &$ 21.28 \pm 0.03 $&$ 1.66 \pm 0.05 $&$145 \pm 2 $&$ 0.25 \pm 0.01$\\
$1\farcm65$ &$ 21.61 \pm 0.03 $&$ 1.65 \pm 0.05 $&$146 \pm 2 $&$ 0.25 \pm 0.01$\\
$2\farcm00$ &$ 21.94 \pm 0.03 $&$ 1.62 \pm 0.05 $&$145 \pm 2 $&$ 0.27 \pm 0.01$\\
$2\farcm20$ &$ 22.19 \pm 0.03 $&$ 1.60 \pm 0.05 $&$145 \pm 2 $&$ 0.26 \pm 0.01$\\
$2\farcm56$ &$ 22.56 \pm 0.03 $&$ 1.59 \pm 0.05 $&$143 \pm 2 $&$ 0.26 \pm 0.01$\\
$3\farcm10$ &$ 22.92 \pm 0.03 $&$ 1.59 \pm 0.05 $&$144 \pm 2 $&$ 0.28 \pm 0.01$\\
$3\farcm41$ &$ 23.14 \pm 0.03 $&$ 1.59 \pm 0.06 $&$144 \pm 2 $&$ 0.28 \pm 0.01$\\
$4\farcm14$ &$ 23.50 \pm 0.03 $&$ 1.59 \pm 0.07 $&$146 \pm 2 $&$ 0.33 \pm 0.01$\\
$4\farcm54$ &$ 23.80 \pm 0.03 $&$ 1.53 \pm 0.10 $&$146 \pm 2 $&$ 0.33 \pm 0.01$\\
$5\farcm01$ &$ 24.07 \pm 0.03 $&$ 1.53 \pm 0.10 $&$146 \pm 2 $&$ 0.33 \pm 0.01$\\
$5\farcm49$ &$ 24.28 \pm 0.03 $&$ 1.50 \pm 0.12 $&$145 \pm 2 $&$ 0.35 \pm 0.01$\\
$6\farcm04$ &$ 24.63 \pm 0.03 $&                &$144 \pm 2 $&$ 0.35 \pm 0.01$\\
$6\farcm66$ &$ 24.92 \pm 0.03 $&                &$144 \pm 2 $&$ 0.35 \pm 0.01$\\
$7\farcm31$ &$ 25.19 \pm 0.04 $&                &$147  \pm2 $&$ 0.35 \pm 0.01$\\
$8\farcm04$ &$ 25.41 \pm 0.05 $&                &$157  \pm3 $&$ 0.37 \pm 0.02$\\
$8\farcm86$ &$ 25.49 \pm 0.05 $&                &$156  \pm4 $&$ 0.43 \pm 0.01$\\
$9\farcm73$ &$ 26.01 \pm 0.06 $&                &$156  \pm4 $&$ 0.39 \pm 0.02$\\\hline
\end{tabular}
\caption{The galaxy luminosity profile, color profile, position angle (PA) and
        ellipticity ($\epsilon$) is tabulated. The errors on the position angle and
        ellipticity are those given by the task {\it ellipse}. The luminosity and
        color uncertainties include calibration errors as well.}
\label{tab:galprof}
\end{table}

\begin{table}[h]
\begin{tabular}{cccc}
\hline
\hline
galaxy & $\mathrm{M}_\mathrm{V}$ & $\alpha_\mathrm{GC}$ &Lit \\\hline
NGC\,524 &$ -21.90 \pm 0.22 $&$-1.71 \pm 0.12$ & KP97\\
NGC\,708 &$ -22.41 \pm 0.32$&$-1.00 \pm 0.11$  & OH02\\
NGC\,720 &$ -22.07 \pm 0.18 $&$-2.2 \pm 0.1 $ &  KP96\\
NGC\,1052&$ -21.09 \pm 0.30$&$ -2.26 \pm 0.27$ & KP97\\
NGC\,1316&$ -23.43 \pm 0.19$&$ -2.04 \pm 0.20$ & Gea01\\
NGC\,1374&$ -20.53 \pm 0.15$&$ -1.80 \pm 0.20$ & KP97\\
NGC\,1380&$ -21.40 \pm 0.18$&$ -1.60 \pm 0.10$ & KP97b\\
NGC\,1399&$ -22.48 \pm 0.19$&$ -1.40 \pm 0.15$ & Dea03a\\
NGC\,1404&$ -21.67 \pm 0.23$&$ -2.00 \pm 0.10$ & KP97\\
NGC\,1427&$ -21.03 \pm 0.26$&$ -2.00 \pm 0.20$ & Fea01\\
NGC\,1700&$ -22.54 \pm 0.25$&$ -1.10 \pm 0.07$ & Bea00\\
NGC\,3115&$ -21.23 \pm 0.10$&$ -1.84 \pm 0.23$ & KP97\\
NGC\,3258&$ -21.45 \pm 0.30$&$ -1.60 \pm 0.20$ & Dea03b\\
NGC\,3268&$ -21.68 \pm 0.32$&$ -1.80 \pm 0.20$ & Dea03b\\
NGC\,3379&$ -20.93 \pm 0.11$&$ -1.73 \pm 0.15$ & RZ03$^1$\\
NGC\,4278&$ -21.01 \pm 0.24$&$ -1.85 \pm 0.20$ & KP97\\
NGC\,4365&$ -22.06 \pm 0.18$&$ -1.15 \pm 0.25$ & KP97\\
NGC\,4406&$ -22.41 \pm 0.14$&$ -1.24 \pm 0.05$ & RZ03\\
NGC\,4472&$ -23.21 \pm 0.10$&$ -1.28 \pm 0.04$ & RZ01\\
NGC\,4486&$ -22.54 \pm 0.16$&$ -1.61 \pm 0.08$ & KP97\\
NGC\,4594&$ -22.14 \pm 0.18$&$ -1.85 \pm 0.07$ & RZ03\\
NGC\,4636&$ -21.78 \pm 0.24$&$ -1.30 \pm 0.10$ & this paper\\
NGC\,4697&$ -21.30 \pm 0.18$&$ -1.90 \pm 0.20$ & KP97\\
NGC\,7014&$ -21.82 \pm 0.34$&$ -1.71 \pm 0.11$ & OH02\\
NGC\,5193&$ -21.99         $&$ -1.38 \pm 0.07$ & OH02\\
IC\,5193&$  -23.49         $&$ -1.02 \pm 0.08$ & OH02\\\hline
\end{tabular}
\caption{Total absolute luminosities of the galaxies (taken from
        either the Caon et al. \cite{caon94} - transformed into V - or
        from the RC3.9) and the
        power-law exponent of its GCS that are plotted as filled circles
        in the lower panel of Fig.\ref{dirsch.fig15}. KP97 - Kissler-Patig
        (\cite{kisslerpatig97}), KP97b - Kissler-Patig et al. (\cite{kisslerpatig97b}),
        KP96 - Kissler-Patig et al. (\cite{kisslerpatig96}),
        RZ01 - Rhode\,\&\,Zepf (\cite{rhode01}), RZ03 - Rhode\,\&\,Zepf (\cite{rhode03}),
        Fea91 - Forte et al. (\cite{forte01}), Dea03a - Dirsch et al. (\cite{dirsch03a}),
        Dea03b - Dirsch et al. (\cite{dirsch03b}), Gea01 - Gomez et al. (\cite{gomez01}),
        Bea00 - Brown et al. (\cite{brown00}), OH02 - Okon\,\&\,Harris (\cite{okon02}). For
        the last two galaxies no SBF distance is available and thus the distance used in
        the paper on the GCS was assumed. $^1$ We fitted the GC density profile using
        the data points of RZ03 with radial distances smaller than 11\arcmin.}
\label{tab:mv_alpha1}
\end{table}

\begin{table}[h]
\begin{tabular}{cccc}
\hline
\hline
galaxy & $\mathrm{M}_\mathrm{V}$ & $\alpha_\mathrm{gal}$ & $\alpha_\mathrm{gal}-\alpha_\mathrm{GC}$ \\\hline
NGC\,1336 &$ -19.09 \pm 0.21$&$ -2.53 \pm 0.09$ &\\
NGC\,1351 &$ -20.30 \pm 0.16$&$ -2.31 \pm 0.04$ &\\
NGC\,1373 &$ -18.62 \pm 0.47$&$ -2.74 \pm 0.07$ &\\
NGC\,1375 &$ -19.36 \pm 0.13$&$ -2.58 \pm 0.02$ &\\
NGC\,1379 &$ -20.47 \pm 0.15$&$ -2.58 \pm 0.09$ &\\
NGC\,1380 &$ -21.40 \pm 0.18$&$ -2.01 \pm 0.04$ & $0.39\pm0.11$\\
NGC\,1380A&$ -18.70 \pm 0.29$&$ -2.65 \pm 0.04$ &\\
NGC\,1381 &$ -19.95 \pm 0.21$&$ -2.61 \pm 0.05$ &\\
NGC\,1399 &$ -22.48 \pm 0.16$&$ -1.67 \pm 0.01$ & $0.27\pm0.15$\\
NGC\,1404 &$ -21.67 \pm 0.19$&$ -2.28 \pm 0.05$ & $0.28\pm0.10$\\
NGC\,1419 &$ -18.85 \pm 0.24$&$ -2.85 \pm 0.03$ &\\
NGC\,1427 &$ -21.03 \pm 0.24$&$ -2.22 \pm 0.04$ & $0.22\pm0.20$\\
NGC\,4261 &$ -22.47 \pm 0.19$&$ -1.93 \pm 0.03$ &\\
NGC\,4339 &$ -19.67 \pm 0.18$&$ -2.16 \pm 0.08$ &\\
NGC\,4472 &$ -23.21 \pm 0.10$&$ -1.60 \pm 0.03$ & $0.32\pm0.06$\\
NGC\,4600 &$ -17.04 \pm 0.22$&$ -2.40 \pm 0.07$ &\\
NGC\,4636 &$ -21.78 \pm 0.13$&$ -1.62 \pm 0.05$ & $0.32\pm0.21$\\\hline
\end{tabular}
\caption{Total absolute luminosities of the host galaxies 
	(taken from Caon et al. \cite{caon94}) and the fitted
        power-law exponents of their luminosity profiles that are plotted
        as filled circles in the upper panel of Fig.\ref{dirsch.fig15}.}
\label{tab:mv_alpha2}
\end{table}

\section*{Acknowledgments}
TR and BD gratefully acknowledge support from the Chilean Center for Astrophysics FONDAP No.
15010003. YS gratefully acknowledges funding support from the DAAD (German Academic Exchange 
Service). We thank the anonymous referee for her/his careful reading and helpful comments.

\section*{References}
\begin{quote}

\bibitem[1992]{ashman92} Ashman, K.M., Zepf, S.E., 1992,
        ApJ, 384, 50
\bibitem[1995]{ashman95} Ashman, K.M., Conti,A., Zepf, S.E., 1995,
        AJ, 110, 1164 
\bibitem[1998]{ashman98} Ashman, K.M., Zepf, S.E., 1998,
	Globular cluster systems, Cambridge astrophysics series, Cambridge 
\bibitem[1997]{athana97} Athanassoula, E., Makino, J., Bosma, A. 1997,
        MNRAS 286, 825
\bibitem[1994]{awaki94} Awaki H., Mushotzky R., Tsuru T., Fabian A.C., 1994,
	PASJ, 46, L65
\bibitem[2002]{beasley02} Beasley, M.A., Baugh, C.M., Ducan, A.F. et al., 2002,
        MNRAS 333, 383
\bibitem[1978]{bottinelli78} Bottinelli, L., Gouguenheim, L., 1978,
	A\&A, 64, L3
\bibitem[2000]{brown00} Brown, R.J.M., Forbes D.A., Kissler-Patig, M., Brodie, J.P., 2000,
	MNRAS, 317, 406
\bibitem[1982]{burstein82} Burstein, D., Heiles, C., 1982, 
        AJ, 87, 1165
\bibitem[2003]{cantiello03} Cantiello M., Raimondo G., Brocato E., Capaccioli M. 2003,
        AJ 125, 2783
\bibitem[1994]{caon94} Caon, N., Capaccioli, M., D'Onofrio, M., 1994,
        A\&AS, 106, 199
\bibitem[2002]{cote02} C\^ot\'e, P., West, M.J., Marzke, R.O., 2002,
        ApJ, 567, 853
\bibitem[2003a]{dirsch03a} Dirsch, B., Richtler, T., Geisler, T., et al., 2003a,
	AJ, 125, 1908
\bibitem[2003b]{dirsch03b} Dirsch, B., Richtler, T., Bassino, L.P., 2003b,
	A\&A, 408, 929
\bibitem[1999]{elmegreen99} Elmegreen, B. G., 1999,
	Ap\&SS, 269, 469
\bibitem[1997]{forbes97} Forbes, D.A., Brodie, J.P., Grillmair, C.J., 1997,
        AJ 113, 1652
\bibitem[2001]{forte01} Forte, J.C., Geisler, D., Ostrov, P.G., 2001,
	AJ, 121, 1992
\bibitem[1978]{gallagher78} Gallagher, J.S., 1978,
	IAUS, 77, 54
\bibitem[1996]{geisler96a} Geisler, D., 1996,
        AJ, 111, 480
\bibitem[1939] {giclas39} Giclas, H.L., 1939,
	PASP, 51, 166
\bibitem[2001]{gomez01} G\'omez, M., Richtler, T., Infante, L., Drenkhahn, G., 2001,
	A\&A, 371, 875
\bibitem[2001]{goudfrooij01} Goudfrooij, P., Mack, J., Kissler-Patig, M., et al., 2001,
	MNRAS, 322, 643
\bibitem[1977]{hanes77} Hanes, D.A., 1977,
	MNRAS, 180, 309
\bibitem[1977]{harris77} Harris, H.C., Canterna, R., 1977,
	AJ, 82, 798
\bibitem[1981]{harris81} Harris, W.E., van den Bergh, S., 1981,
	AJ, 86, 1627
\bibitem[1996]{harris96} Harris, W.E., 1996, 
	AJ, 112, 1487
\bibitem[2001]{harris01} Harris, W.E., 2001, in ''Star Clusters``, 
	Saas-Fee Advanced Course 28, Lecture Notes 1998, eds. L. Labhardt and B. Binggeli, 
        Springer-Verlag, Berlin, p.223
\bibitem[2002]{harris02} Harris, W.E., Harris, G.L.H., 2002,
	AJ, 123, 3108
\bibitem[1996]{hilker96} Hilker, M., Kissler-Patig, M., 1996,
	A\&A, 314, 357
\bibitem[2002]{idiart02} Idiart, T.P., Michard, R., de\,Freitas\,Pacheco, J.A., 2002,
	A\&A, 383, 30
\bibitem[1987]{jaffe87} Jaffe, W., 1987,
	in ''Structure and Dynamics of Elliptical Galaxies'', IAU Symp. 127, ed. T. de Zeeuw,
	D. Reidel, Dordrecht, p.511
\bibitem[2002]{jones02} Jones, C., Forman, W., Vikhlinin, A. et al. 2002,
	ApJ, 567, L115
\bibitem[1997]{kisslerpatig97} Kissler-Patig, M., 1997,
	A\&A, 319, 83
\bibitem[1997b]{kisslerpatig97b} Kissler-Patig, M., Richtler, T., Storm, J, della Valle, M., 1997b,
	A\&A, 327, 503
\bibitem[1996]{kisslerpatig96} Kissler-Patig, M., Richtler, T., Hilker, M., 1996,
	A\&A, 308, 704
\bibitem[1994]{kissler94} Kissler, M., Richtler, T., Held, E.V. et al., 1994,
	A\&A, 287, 463
\bibitem[1978]{knapp78} Knapp, G.R., Faber, S.M., Gallagher, J.S., 1978,
	AJ, 83, 11
\bibitem[1989]{kormendy89} Kormendy, J., Djorgovski, S., 1989, 
	ARA\&A, 27, 235
\bibitem[1983]{krishna83} Krishna Kumar, C., Thonnard, N., 1983,
	AJ, 88, 260
\bibitem[1999]{kundu99} Kundu, A., Whitmore, B.C., Sparks, W.B. et al., 1999,
	ApJ, 513, 733
\bibitem[2003]{larsen03} Larsen, S.S., Brodie, J.P., Beasley M.A. et al., 2003,
        ApJ, 585, 767
\bibitem[2001]{larsen01} Larsen, S.S., Brodie, J.P., Huchra, J.P. et al., 2001,
        AJ, 121, 2974
\bibitem[2000]{larsen00} Larsen, S.S., Richtler T., 2000,
	A\&A, 354, 836
\bibitem[2000]{leonardi00} Leonardi, A. J., Worthey, G., 2000,
	ApJ, 534, 650
\bibitem[1995]{lauer95} Lauer, T.R., Ajhar, E.A., Byun, Y. et al., 1995,
        AJ 110,  2622
\bibitem[2003]{loewenstein03} Loewenstein, M., Mushotzky, F., 2003,
	Nuclear Physics B Proc. Suppl. 124, 91
\bibitem[1998]{matsushita98} Matsushita, K., Makishima, K., Ikebe, Y. et al., 1998,
	ApJ, 499, L13
\bibitem[1999]{mclaughlin99} McLaughlin, D., 1999, 
        AJ, 117, 2398
\bibitem[1996]{mcmillan96} Mc\,Millan, R.J., Ciardullo, R., 1996,
	ApJ, 473, 707
\bibitem[1994]{merritt94} Merritt, D., Tremblay, B., 1994,
        AJ 111, 2243
\bibitem[1996]{minniti96} Minniti, D., Alsonso, M.V., Goudfrooij, P. et al., 1996,
	ApJ, 467, 221	
\bibitem[1999]{mulchaey99} Mulchaey J.S., Zabludoff A.I., 1999,
        ApJ, 514, 133
\bibitem[1994]{mushotzky94} Mushotzky, R.F., Loewenstein, M., Awaki, H. et al., 1994,
	ApJ, 436, L79 
\bibitem[2002]{okon02} Okon, W.M.M., Harris, W.E., 2002,
	ApJ, 567, 294
\bibitem[1989]{ostriker89} Ostriker, J. P., Binney, J., Saha, P., 1989,
	MNRAS, 241, 849
\bibitem[1998]{ostrov98} Ostrov, P.G., Forte, J.C., Geisler, D., 1998,
        AJ, 116, 2854
\bibitem[2003]{otho03} Otho, A., Kawano, N., Fukazawa, Y., 2003,
	PASJ, 55, 819
\bibitem[1998]{prugniel98}  Prugniel, P., Heraudeau, P., 1998,
        A\&AS, 128, 299
\bibitem[1999]{puzia99} Puzia, T.H., Kissler-Patig, M., Brodie, J.P., Huchra, J.P., 1999,
	AJ, 118, 2734
\bibitem[2001]{rhode01} Rhode, K.L., Zepf, S.E., 2001,
	AJ, 121, 210
\bibitem[2003]{rhode03} Rhode, K.L., Zepf, S.E., 2003,
	astro-ph/0310277
\bibitem[2003]{richtler03} Richtler, T., 2003,
	astro-ph/0304318
\bibitem[1985]{rieke85} Rieke G.H., Lebovsky M.J., 1985,
	ApJ, 288, 618
\bibitem[1998]{schlegel98} Schlegel, D., Finkbeiner, D., Davis, M., 1998,
        ApJ, 500, 525
\bibitem[1986]{schombert86} Schombert, J.M., 1986,
        ApJS, 60, 603
\bibitem[1992]{secker92} Secker, J., 1992,
	AJ, 104, 1472
\bibitem[1993]{secker93} Secker, J., Harris, W.E., 1993,
	AJ, 105, 1358
\bibitem[2004]{stocke04} Stocke J.T, Keeney B.A., Lewis A.D., Epps H.W., Schild R.E. 2004,
        AJ, 127, 1336
\bibitem[2003]{temi03} Temi, P., Mathews, W.G., Brighenti, F., Bregman, J.D., 2003,
	ApJ, 585, L121
\bibitem[2001]{tonry01} Tonry, J.L., Dressler, A.D., Blakeslee, J.P., 2001,
        AJ, 546, 681
\bibitem[1994]{trinchieri94} Trinchieri, G., Kim, D.-W., Fabbiano, G., Canizares, C.R.C., 1994,
	ApJ, 428, 555
\bibitem[1999]{vikh99} Vikhlinin A., McNamara B.R., Hornstrup A. et al. 1999,
        ApJ 520, L1
\bibitem[1987]{white87} White, S.D.M. 1987,
	in ''Structure and Dynamics of Elliptical Galaxies'', IAU Symp. 127, ed. T. de Zeeuw,
	D. Reidel, Dordrecht, p.263
\bibitem[1999]{whitmore99} Whitmore, B.C., Zhang, Q., Leitherer, C., at al., 1999,
	AJ 118, 1551 
\bibitem[2002]{whitmore02} Whitmore, B.C., Schweizer, F., Kundu, A., Miller, B.W., 2002,
	AJ, 124, 147
\bibitem[2002]{xu02} Xu, H., Kahn, S.M., Peterson, J.R. et al., 2002,
	ApJ, 579, 600
\bibitem[1996]{yoshii96} Yoshii, Y., Tsujimoto, T., Nomoto, K., 1996,
	ApJ, 462, 266
\bibitem[1995]{yungelson95} Yungelson, L., Livio, M., Tutukov, A., Kenyon, S.J., 1995,
	ApJ, 447, 656
\end{quote}

\end{document}